\documentclass[prb,aps,twocolumn,showpacs]{revtex4-1}

\usepackage{graphicx}
\usepackage{latexsym}
\usepackage{amsmath}
\usepackage{bm}

\newcommand{\be}{\begin{eqnarray}}
\newcommand{\ee}{\end{eqnarray}}
\newcommand{\nn}{\nonumber\\}

\newcommand{\la}{\langle}
\newcommand{\ra}{\rangle}

\begin{document}

\title{Finite-temperature phase diagram of 
two-component bosons in a cubic optical lattice:
Three-dimensional $t$-$J$ model of hard-core bosons}

\date{\today}

\author{Yuki Nakano$^1$} 
\author{Takumi Ishima$^2$} 
\author{Naohiro Kobayashi$^2$} 
\author{Takahiro Yamamoto$^1$}
\author{Ikuo Ichinose$^2$}
\author{Tetsuo Matsui$^1$}
\affiliation{%
$^1$Department of Physics, Kinki University, 
Higashi-Osaka, 577-8502 Japan
}%
\affiliation{%
$^2$Department of Applied Physics, Nagoya Institute of Technology,
Nagoya, 466-8555 Japan}

\begin{abstract}
We study the three-dimensional bosonic $t$-$J$  model, i.e.,
the $t$-$J$ model of ``bosonic electrons", at finite temperatures.
This model describes the $s={1 \over 2}$ Heisenberg spin model with 
the anisotropic exchange coupling $J_{\bot}=-\alpha J_z$ and 
doped {\it bosonic} holes, which is an effective system
of the Bose-Hubbard model with strong repulsions.
The bosonic ``electron" operator $B_{r\sigma}$ at the site $r$ 
with a two-component (pseudo-)spin $\sigma (=1,2)$ 
is treated as  a hard-core boson operator, 
and represented by a composite of two slave particles;
a ``spinon" described by a Schwinger boson (CP$^1$ boson) $z_{r\sigma}$
and a ``holon" described by a hard-core-boson field $\phi_r$
as $B_{r\sigma}=\phi^\dag_r z_{r\sigma}$. 
By means of Monte Carlo simulations, 
we study its finite-temperature phase structure including the $\alpha$ 
dependence, the possible  phenomena like appearance of checkerboard
long-range order, super-counterflow, superfluid, and phase separation, etc.
The obtained results may be taken as predictions about experiments 
of two-component cold bosonic atoms in the cubic optical lattice.
\end{abstract}

\pacs{67.85.Hj, 75.10.-b, 03.75.Nt}

\maketitle

\section{Introduction}
\setcounter{equation}{0} 

Cold-atomic systems are one of the most intensively studied topics not only in
atomic physics but also in condensed matter physics in these days.
In particular, cold atoms put on an optical lattice (OL) may be used 
as a ``simulator" to study  certain canonical models of
strongly-correlated electron systems\cite{optical}.
For systems in the OL, interactions between atoms, 
dimensionality of system, etc. are highly controllable,
and effects of impurities are strongly suppressed.
Therefore,  cold atomic systems in the OL 
are sometimes called final simulators.
Among them, systems of double-species atoms are quite interesting
from the view point of the high-temperature ($T$) superconductivity (SC). 
Investigation of these atomic systems is expected to give an important insight
into mechanism of SC in systems in which only repulsive 
interactions between particles exist.

In this paper, we shall study the $t$-$J$ model 
of hard-core bosons in the cubic lattice.
There are (at least) two versions of the bosonic $t$-$J$ model\cite{BtJ1}.
In the previous paper\cite{BtJ1}, we studied one version that 
is a bosonic counterpart of the original fermionic
$t$-$J$ model and respects the SU(2) spin symmetry.
On the other hand,
in the present paper, we shall consider the second version that is an effective
model of the two-band Bose-Hubbard model with strong repulsions
and the total filling factor not exceeding unity\cite{BtJ2}.
Relation and differences between these two versions of the bosonic 
$t$-$J$ model were explained in the previous paper\cite{BtJ1}. 
Obtained results for the second version of the $t$-$J$ model in the present paper
can be regarded as predictions
about the system of bosonic atoms of two-species.
Related Hubbard model at commensurate fillings
has been studied in e.g., Ref.\cite{altman}
by the mean-field-theory (MFT) type approximation, and its one-dimensional
counterpart by the Tomonaga-Luttinger liquid theory in Ref.\cite{Hu}.
In the present paper, we shall study the three-dimensional (3D) 
$t$-$J$ model at fractional fillings
mostly by means of the Monte-Carlo (MC) simulations.
Results are compared with the ones obtained previously.

The paper is organized as follows.
In Sect.2 we explain the model and  its basic properties.
We also study it by MFT briefly.
In Sect.3 we present the results of MC simulations.
Section 4 is devoted for conclusions and discussions.
 
\section{Model}
\setcounter{equation}{0} 

\subsection{The $t$-$J$ model}

The $t$-$J$ model is derived from the 
Bose-Hubbard model\cite{BHM} whose Hamiltonian is given as
\begin{eqnarray}
H_{\rm Hub}&=&\sum_r\Big[-t\sum_{i=1}^3 (a^\dagger_{r+i}a_r+b^\dagger_{r+i}b_r
+\mbox{h.c.})
+Un_{ar}n_{br}   \nonumber  \\
&&+{V \over 2}\sum_{\lambda=a,b} n_{\lambda r}(n_{\lambda r}-1)
-\mu_{c}\sum_{\lambda=a,b}n_{\lambda r}\Big],
\label{HHu}
\end{eqnarray}
where $r$ denotes site of the cubic lattice, $i(=1,2,3)$ is the 
unit vector in the $i$-th direction (it also denotes the
direction index), and $a_r$ and $b_r$ are boson annihilation operators.
$n_\lambda$ is the number operator of the boson $\lambda$, and therefore
$U$ and $V$ are inter- and intra-species interactions, respectively.
This $H_{\rm Hub}$ describes the system of two-species of
cold bosonic atoms in a cubic OL.
From Eq.(\ref{HHu}), it is obvious that $a$ and $b$ atoms have the same
hopping amplitude and the same density $\rho_a=\rho_b$ in the present system.
Recently studied $^{85}$Rb -$^{87}$Rb atomic system\cite{2BEC} is a typical example
described by this Hamiltonian.

Some related models to $H_{\rm Hub}$ in Eq.(\ref{HHu}) have been studied so far.
In the present paper, we consider the specific case  such that
$t \ll U,V$
and {\em the total number of bosons at each site is not exceeding unity}
$(0 \leq n_{ar}+n_{br}\leq 1$).
It is obvious that the model in the above parameter region is closely related with the 
high-$T_{\rm c}$ materials and therefore it is expected that study on it
gives rise to an important insight into the physical properties of 
the high-$T_{\rm c}$ materials.
It should be remarked that at present the properties of the fermionic 
$t$-$J$ model
are poorly understood in spite of the quite intensive studies on it for more
than two decades.
This fact mainly stems from the difficulties of numerical study on fermionic systems.

The effective Hamiltonian in the large on-site repulsion limit can be 
derived by the standard methods of expansion in powers of $t/U,
t/V$\cite{effectiveH} as follows;
\begin{eqnarray}
H_{tJ}&=&-t\sum_{r,i=1}^3 (a^\dagger_{r+i}a_r
+b^\dagger_{r+i}b_r+\mbox{h.c.})
+J_z\sum_{r,i}S^z_{r+i}S^z_r  \nonumber  \\
&& \hspace{-1cm} -J_{\bot}\sum_{r,i}(S^x_{r+i}S^x_r+S^y_{r+i}S^y_r)  
-\bar{\mu}_c\sum_{r}(1-n_{ar}-n_{br}),\nn
\label{HtJ}
\end{eqnarray}
where the SU(2) pseudo-spin operator is given as
 $\vec{S}_r={1 \over 2}B^\dagger_r\vec{\sigma}B_r$ with
$B_r=(a_r,b_r)^t$ ($\vec{\sigma}$ is the Pauli spin matrices).
The exchange couplings are
\begin{equation}
J_z={4t^2 \over U}-{4t^2 \over V}+\cdots, \
J_{\bot}={2t^2 \over U}+\cdots,\ J_\bot=-\alpha J_z,
\label{Js}
\end{equation}
and $\bar{\mu}_c$ is the chemical potential of holes.
In the following discussion, we shall treat $t,\; J_z$ and $J_{\bot}$, 
hence $\alpha$,  as free parameters, and study the $t$-$J$ model (\ref{HtJ}). 
After obtaining the critical couplings etc, we shall return to the
expression (\ref{Js}).

\subsection{Physical-state condition: Double-CP$^1$ representation}

In the system of $H_{tJ}$ in Eq.(\ref{HtJ}), 
a physical state at each site $r$
is expanded by three orthogonal basis state vectors 
$\{|0\rangle, |a\rangle=a^\dagger_r|0\rangle, |b\rangle=
b^\dagger_r|0\rangle\}\ (a_r|0\rangle = b_r|0\rangle=0)$.
In order to express this constrained Hilbert space faithfully, 
we use the following slave-particle representation,
\begin{eqnarray}
&& a_r=\phi^\dagger_r c_{r1}, \;\;\; 
b_r=\phi^\dagger_r c_{r2},  \label{slave}  \\
&& \Big(\phi^\dagger_r\phi_r+c^\dagger_{r1}c_{r1}+c^\dagger_{r2}c_{r2}-1\Big)
|\mbox{phys}\rangle =0,
\label{const}
\end{eqnarray}
where $\phi_r$ is a hard-core boson and $c_{r\sigma}\ (\sigma=1,2)$ 
is an ordinary boson.
The three basis states are expressed in terms of $c_{r\sigma}$ and 
$|\Omega\ra\ (c_{r\sigma}|\Omega\rangle=0)$ as 
\begin{equation}
|0\rangle \leftrightarrow \phi^\dagger_r|\Omega\rangle, \;\;
a^\dagger_r|0\rangle \leftrightarrow c^\dagger_{r1}|\Omega\rangle, \;\;
b^\dagger_r|0\rangle \leftrightarrow c^\dagger_{r2}|\Omega\rangle.
\label{slave2}
\end{equation}

In order to express the local constraint (\ref{const}) in more 
convenient way, we introduce a CP$^1$ boson (Schwinger boson) $z_{r\sigma}$,
\begin{eqnarray}
&& c_{r\sigma}=(1-\phi^\dagger_r\phi_r)z_{r\sigma}, \;\; (\sigma=1,2)  \nonumber \\
&& \Big(\sum_{\sigma=1,2}z^\dagger_{r\sigma}z_{r\sigma}-1\Big)
|\mbox{phys}\rangle_{z} =0.
\label{CP1}
\end{eqnarray}
It is easily verified that Eq.(\ref{const}) is satisfied by Eq.(\ref{CP1}).

The hard-core boson $\phi_r$ itself can be expressed in terms of 
another CP$^1$ boson
$w_{rf}$ as follows\cite{BtJ1},
\begin{equation}
\phi_r=w^\dagger_{r2}w_{r1}, \;\;
\Big(\sum_{f=1,2}w^\dagger_{rf}w_{rf}-1\Big)
|\mbox{phys}\rangle_{w} =0.
\label{w}
\end{equation}
From Eq.(\ref{w}), it is obvious that 
$|0\rangle_\phi=w^\dagger_{r2}|0\rangle_w$ and 
$\phi^\dagger_r|0\rangle_\phi=w^\dagger_{r1}|0\rangle_w$,
where $|0\rangle_\phi (|0\rangle_w)$ is the empty state of 
$\phi_r (w_r)$.
It is straightforward to verify that $\phi_r$ satisfies the mixed commutation
relations of hard-core bosons\cite{BtJ1}.
Then the Hamiltonian $H_{tJ}$ can be expressed in terms
of the two sets of  CP$^1$ bosons $z_{r\sigma}$ and $w_{rf}$.
The partition function 
$Z$ at finite $T$ is given by the path-integral as
\begin{equation}
Z=\int [DwDz]\exp[-\int_0^\beta d\tau
(\bar{z}\dot{z}+\bar{w}\dot{w}+H_{tJ}(\bar{w},w,\bar{z},z))],
\label{ZGCE0}
\end{equation}
where $\tau$ is the imaginary time, 
$\beta=1/(k_{\rm B}T)$ and $H_{tJ}(\bar{w},w,\bar{z},z)$
is obtained from $H_{tJ}$ in (\ref{HtJ}) by substituting 
the double-CP$^1$ representation for $a_r$ and $b_r$.
In the present numerical study, we ignore the $\tau$-dependence
of $z(\tau)$ and $w(\tau)$ and consider the following system,
\begin{equation}
Z'=\int [DwDz]\exp[-\beta H_{tJ}(\bar{w},w,\bar{z},z)],
\label{ZGCE}
\end{equation}
where $z$ and $w$ represent the zero-modes of $z(\tau)$
and $w(\tau)$.
This approximation is justified when we consider system at
sufficiently high temperature.
However, we expect that the system (\ref{ZGCE}) has at least qualitatively
the same phase structure to that of (\ref{ZGCE0}) for $T>0$.
As we discussed in Ref.\cite{BtJ1}, the nonzero-modes of 
$z(\tau)$ and $w(\tau)$ renormalize $H_{tJ}$ and this renormalization
tends to order the system.
Therefore it is expected that
ordered phase found in the system (\ref{ZGCE}) also exists in 
the system (\ref{ZGCE0}).
This expectation was actually verified in some systems\cite{sawa}.

\subsection{Mean-field theory}

Before going into the details of numerical study of Eq.(\ref{ZGCE}),
it is useful to investigate the ground-state properties of the model
by the MFT.
We use a variational wave function of bosons $a_r$ and $b_r$ that has 
a site-factorized form,
\begin{equation}
|\Psi\rangle=\prod_r\Big[\sin {\theta_r \over 2}
\Big(\sin {\chi_r \over 2}a^\dagger_r+\cos {\chi_r \over 2}b^\dagger_r
\Big)+\cos {\theta_r \over 2}\Big] |0\rangle.
\label{vwf}
\end{equation}
Here we assume the sublattice symmetry and put
$\theta_r=\theta$, $\chi_r=\chi_{\rm A(B)} \ [r\in \mbox{A(B)-sublattice}]$.
Then the mean-field energy ${\cal E}_{tJ}$ is given as
\begin{eqnarray}
{{\cal E}_{tJ}\over N_L}&=&
-{t \over 2}\sin^2\theta\Big(\sin{\chi_{\rm A} \over 2}
\sin{\chi_{\rm B} \over 2}+\cos{\chi_{\rm A} \over 2}
\cos{\chi_{\rm B} \over 2}\Big)  \nonumber  \\
&&+{J_z \over 4}\sin^4{\theta \over 2}\cos\chi_{\rm A}\cos\chi_{\rm B}
-{J_{\bot} \over 4}\sin^4{\theta \over 2}\sin\chi_{\rm A}\sin\chi_{\rm B}  \nonumber \\
&&-\bar{\mu}_c\cos^2{\theta \over 2},
\label{EtJ2}
\end{eqnarray}
where $N_L$ is the number of links in the system.
By minimizing ${\cal E}_{tJ}$, we can obtain the MF phase diagram.
The case in which the filling is unity, 
$n_{ra}+n_{rb}=1$, corresponds to $\theta=\pi$
in (\ref{vwf}), and 
\begin{equation}
{{\cal E}_{tJ}\over N_L}\Big|_{\theta=\pi}= 
{J_z \over 4}\cos\chi_{\rm A}\cos\chi_{\rm B}
-{J_{\bot} \over 4}\sin\chi_{\rm A}\sin\chi_{\rm B}.  
\label{EtJ3}
\end{equation}
The lowest-energy state there is easily obtained as 
\begin{eqnarray}
&&\mbox{For} \; J_z>J_{\bot}, \;\; (\chi_{\rm A}, \chi_{\rm B})=(0,\pi), \;\;
\mbox{or} \;\; (\chi_{\rm A}, \chi_{\rm B})=(\pi,0), \nonumber \\
&&\mbox{For}\; J_{\bot}>J_z>-J_{\bot}, \;\; (\chi_{\rm A}, \chi_{\rm B})=
\Big({\pi \over 2},{\pi \over 2}\Big).
\end{eqnarray}
Then for $J_z>J_{\bot}$, the lowest-energy state is the checkerboard state
of particle $a$ and $b$, whereas for $J_{\bot}>J_z$ the state of 
super-counter-flow (SCF)
$\langle a^\dagger_r b_r\rangle \neq 0$ is realized as expected.
The checkerboard state corresponds to an antiferromagnetic (AF) state,
whereas the SCF corresponds to a XY-ferromagnetic state in the magnetism terminology.

\begin{figure}[b]
\begin{center}
\includegraphics[width=7cm]{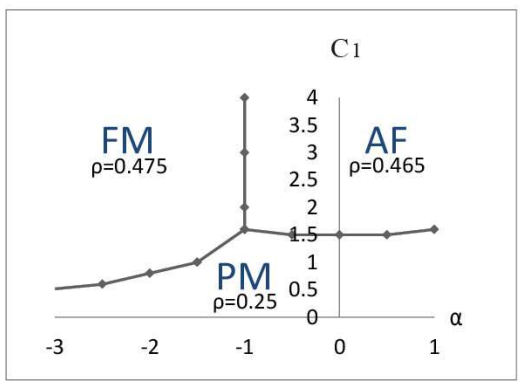}
\caption{
Phase structure for $t=0$ and $\bar{\mu}_c=0$ in the $\alpha-c_1$ plane,
where $\alpha=-J_{\bot}/J_z$ and $c_1=\beta J_z$. 
There are three  phases and
their physical meaning is explained in the text.
Typical value of $\rho=\langle a^\dagger_r a_r \rangle=
\langle b^\dagger_r b_r \rangle$ in each phase is also shown.
}
\label{fig:PD1}
\end{center}
\end{figure}

Doping holes shifts $\theta$ to $\theta<\pi$.
From Eq.(\ref{EtJ2}),
the lowest-energy state is obtained as
\be
&&{\rm For}\ J_{\bot}>J_z, \chi_{\rm A}=\chi_{\rm B}={\pi \over 2},\nn
&&{\rm For}\ J_z>J_{\bot}, \chi_{\rm A}(\chi_{\rm B})=\epsilon,\; 
\chi_{\rm B}(\chi_{\rm A})=\pi-\epsilon,\nn
&&\epsilon\equiv{4t \over J_z-J_{\bot}}\cos^2 {\theta \over 2}
={4t \over J_z-J_{\bot}}(1-n_a-n_b).
\ee
Thus, for $J_{\bot}>J_z$   Bose-Einstein condensation (BEC) 
of both $a$ and $b$ atoms occurs in addition to the SCF.
On the other hand, for $J_z>J_{\bot}$,
superfluidity (SF)
with checkerboard symmetry, so called supersolid (SS),
appears for an arbitrary small but finite value of $t$ and the hole
density\cite{SS}.
However, this result by the MFT is not 
reliable even for the present three-dimensional system because
fluctuations of the
relative phases of $|0\rangle, \; a^\dagger_r|0\rangle$ and
$b^\dagger_r|0\rangle$ have been ignored in the MFT.

In the following section, we shall study the model by means of the MC simulations.
The numerical study gives reliable result for the phase structure
of the model and also details of its critical behavior.

\section{Results of MC simulations}
\setcounter{equation}{0} 

\subsection{Case of $t=0$}

Let us turn to the numerical study\cite{MC}.
We conisder the cubic lattice with its linear size $L$ up to 20
and impose the periodic boundary condition.
In order to find phase transition lines, we calculate the
internal energy $U$ and 
the specific heat $C$ defined as
\be
U=\frac{1}{N}\langle H_{tJ} \rangle,\
C=\frac{1}{N}\langle (H_{tJ}-U)^2\rangle,\ N\equiv L^3.
\ee
Furthermore we calculate various correlation functions to
identify each observed phase.

\begin{figure}[t]
\begin{center}
\includegraphics[width=4.4cm]{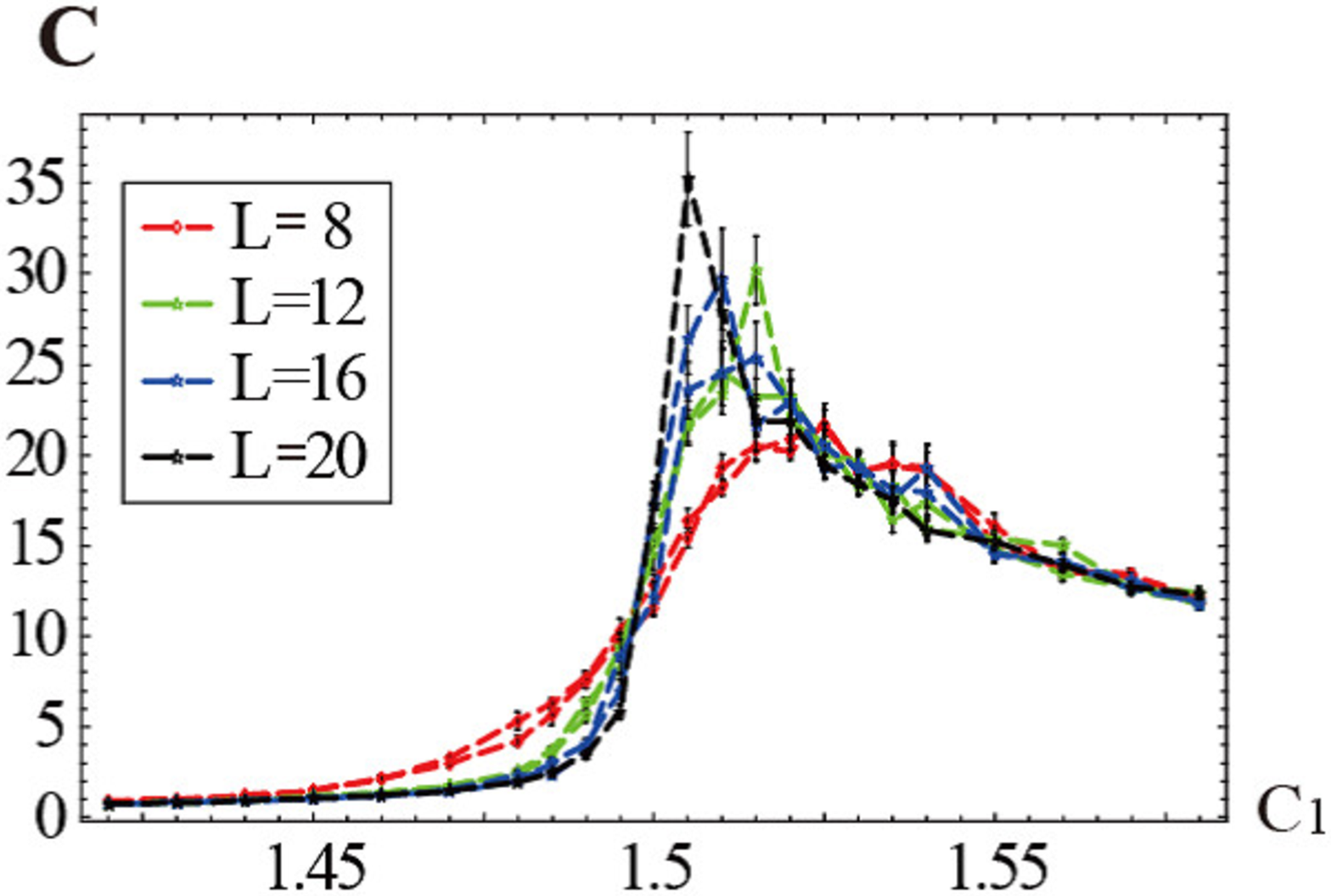}
\hspace{-0.6cm}
\includegraphics[width=4.2cm]{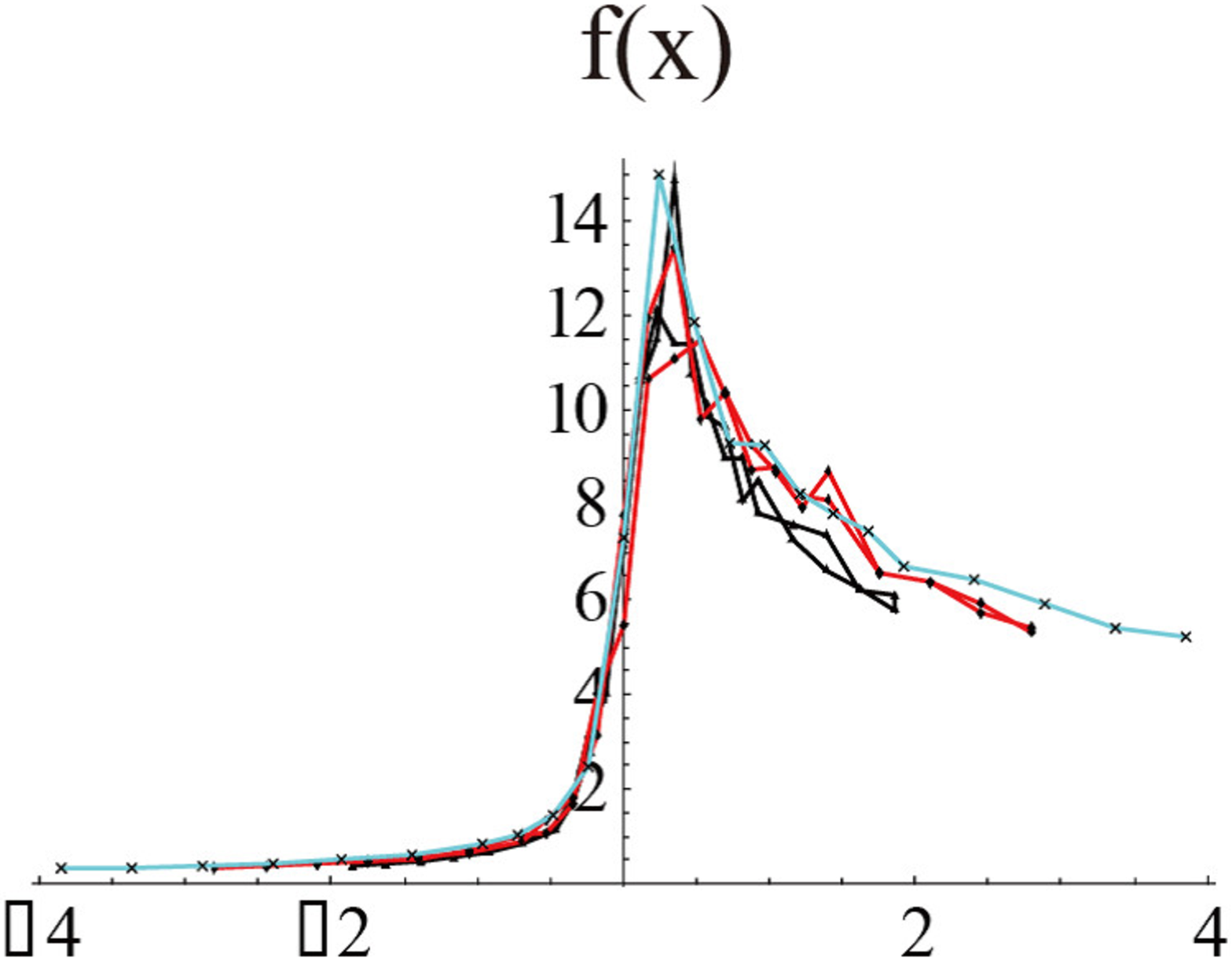}
\caption{
(Left) $C$ at $t=0$ as a function of $c_1$ for $\alpha=-0.5$.
It indicates the existence of a second-order phase transition
(PM $\rightarrow$ AF) at $c_1\simeq 1.51$.
(Right) The scaling function $f(x)$ for $C$ obtained by finite-size scaling (FSS)
hypothesis\cite{FSS}.
The critical exponent of the correlation length is $\nu=0.70$.
}
\label{fig:UC1}
\end{center}
\end{figure}
\begin{figure}[b]
\begin{center}
\hspace{-0.5cm}
\includegraphics[width=4.5cm]{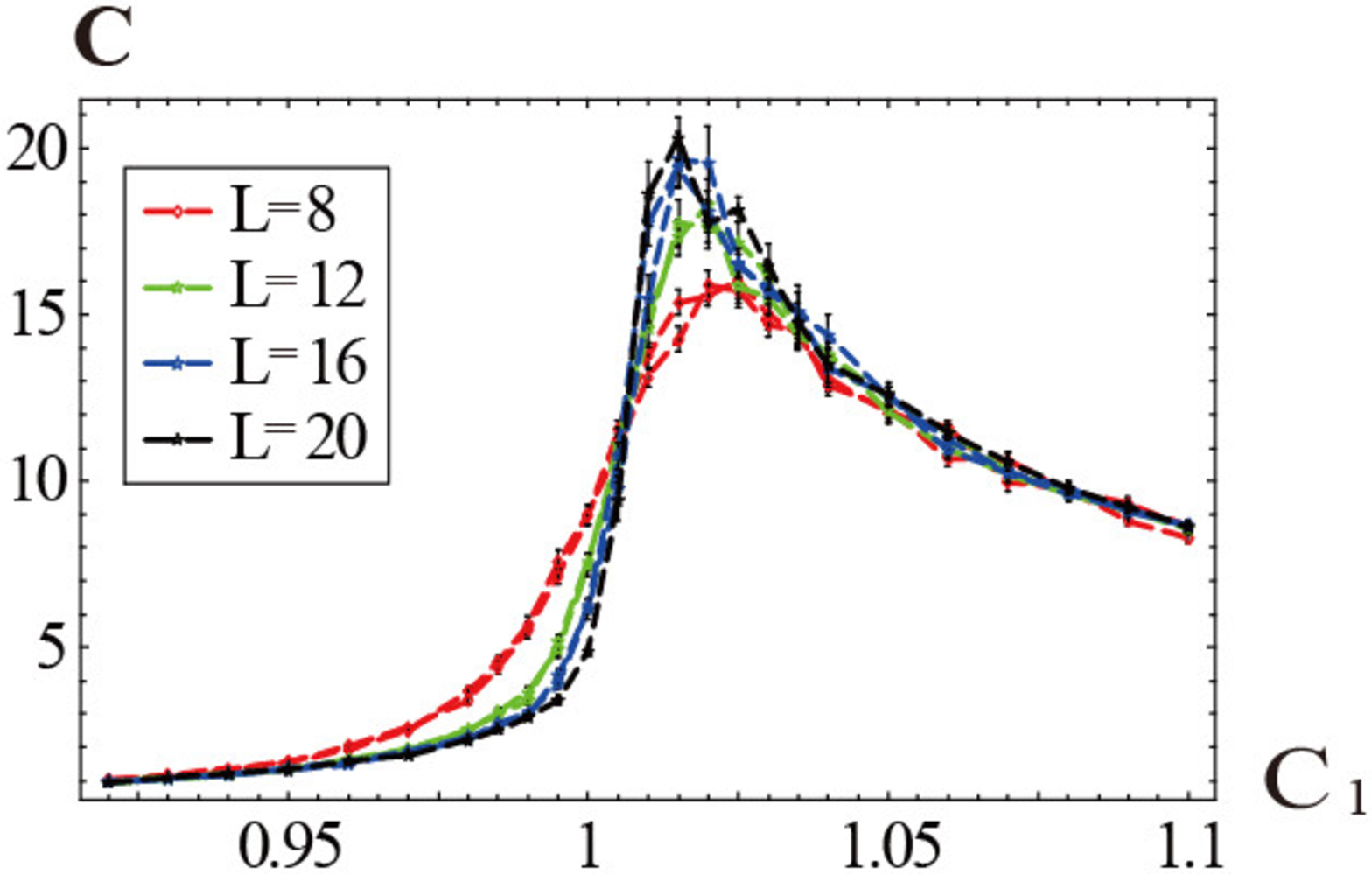}
\includegraphics[width=4cm]{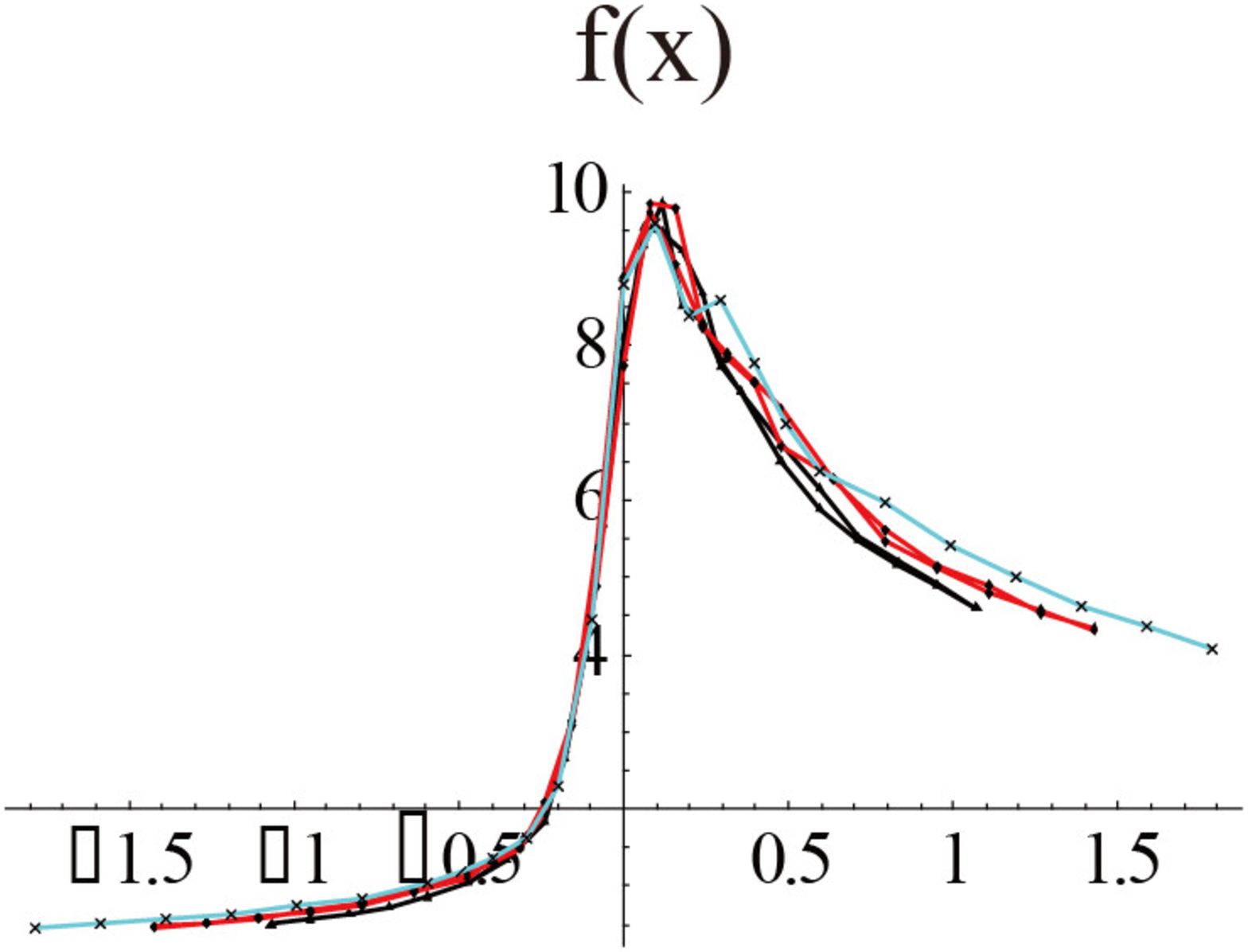}
\caption{
(Left) $C$ as a function of $c_1$ for $\alpha=-1.5$.
Result indicates the existence of second-order phase transition
(PM $\rightarrow$ FM) at $c_1\simeq 1.015$.
(Right) Scaling function $f(x)$.
Critical exponent is obtained as  $\nu=1.0$.
}
\label{fig:UC2}
\end{center}
\end{figure}
\begin{figure}[t]
\begin{center}
\includegraphics[width=4.2cm]{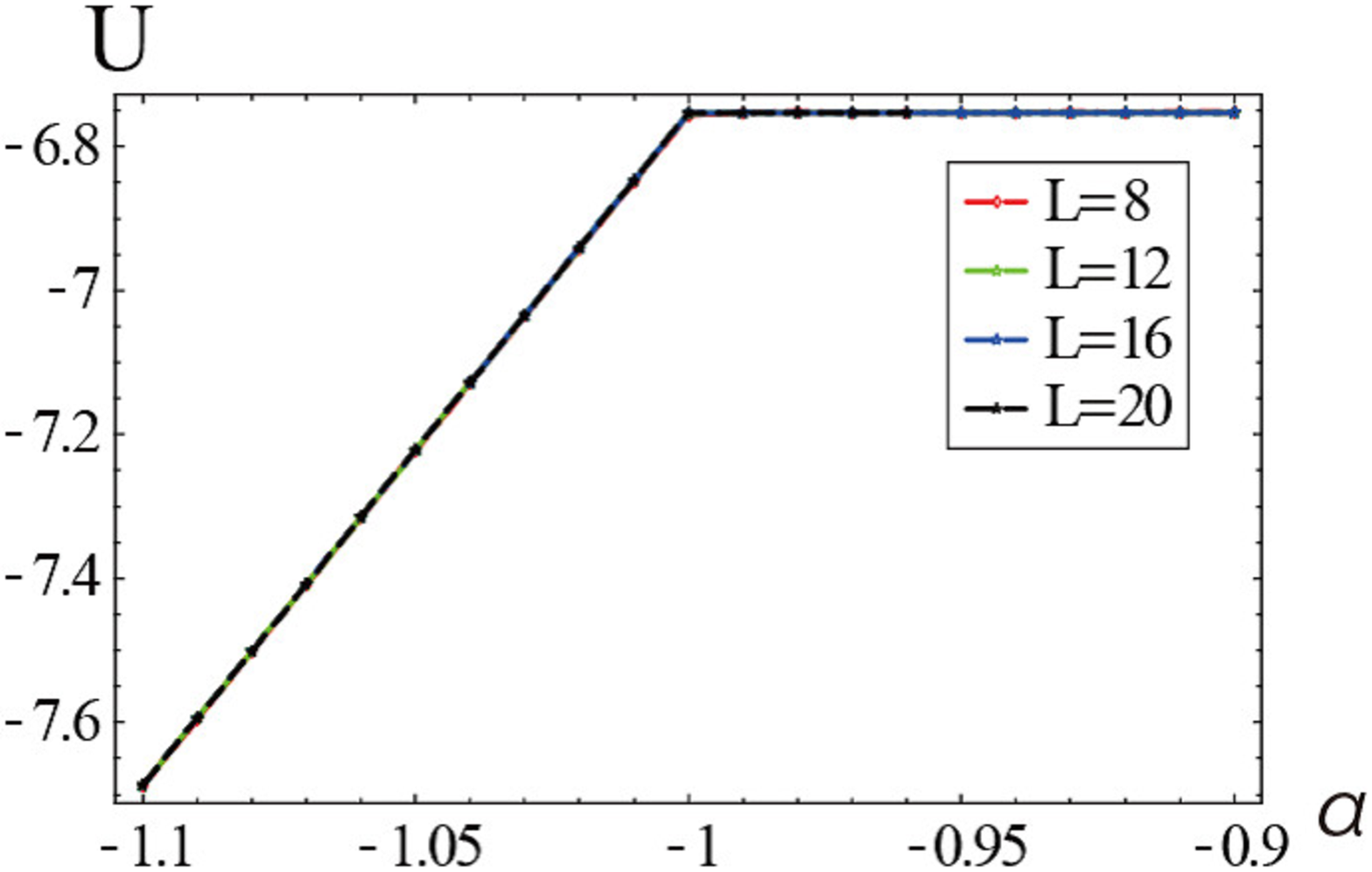}
\includegraphics[width=4.2cm]{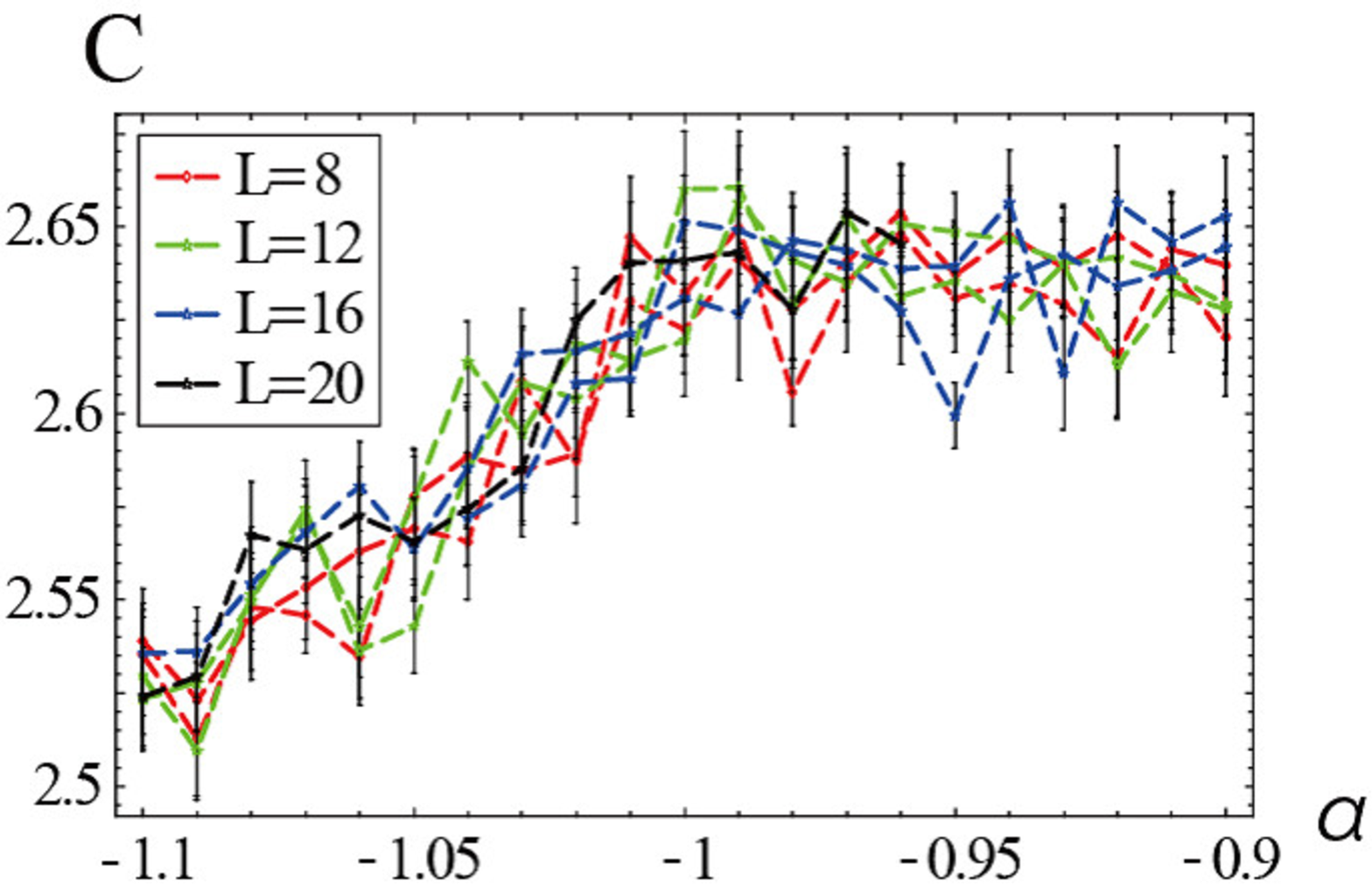}
\caption{
$U$ and $C$ at $t=0$ as a function of $\alpha$ for $c_1=3$.
Phase transition takes place at $\alpha\simeq -1.0$.
}
\label{fig:UC3}
\end{center}
\end{figure}

It is convenient to use the following parameterization, 
\be
\alpha=-\frac{J_{\bot}}{J_z},\ c_1=\beta J_z,\ c_3=\beta t.
\ee
We first consider the case of vanishing hole hopping, i.e., $t=0$.
Phase diagram was obtained for various values of the chemical
potential.
The result for $\bar{\mu}_c=0$
is shown in the $\alpha-c_1$ plane in Fig.\ref{fig:PD1}.
In the following, we shall mostly show results for $\bar{\mu}_c=0$.
Some of calculations of $U$ and $C$, which were used to determine 
the phase boundaries in Fig.\ref{fig:PD1}, are shown in Figs.\ref{fig:UC1},
\ref{fig:UC2} and \ref{fig:UC3}.
In high-$T$ region that corresponds to small $c_1$, the system 
exists in the phase without
any long-range order (LRO), which we call paramagnetic (PM) phase.
As $c_1$ is increased, phase transition to ordered states takes place.
For $\alpha>-1$, AF state with checkerboard configuration
of atoms $a$ and $b$ appears as a result of strong intra-repulsion.
On the other hand, for  $\alpha<-1$, the XY-ferromagnetic state 
appears at low $T$ as a result of strong inter-repulsion.
In the XY-ferromagnetic state, the nonvanishing condensation of
$\langle a^\dagger_r b_r \rangle$ takes place (SCF).
The line $\alpha=-1$, corresponding to $V=2U$, is very specific as 
the symmetry of pseudo-spin degrees of freedom
is enhanced to SU(2) along this line, otherwise the symmetry is
U(1)$\times Z_2$, i.e., a global $(S^x_r-S^y_r)$ rotation and 
$S^z_r \rightarrow  -S^z_r$ reflection.
In the study of ferroelectric materials, the corresponding line is
called morphotropic phase boundary (MPB), and it plays an 
important role\cite{MPB}.
Our calculation in Fig.\ref{fig:UC3} shows that the phase transition at 
$\alpha=-1$ looks neither first order nor second order.
The origin of this peculiar behavior of $U$ and $C$ across the MPB 
is the enhancement of the symmetry at  $\alpha=-1$ as explained.
Turning on the hole-hopping $t$ reduces the symmetry at $\alpha=-1$
down to U(1)$\times Z_2$,
and as a result, the phase transition becomes second-order.
We have studied case of several other values of $\bar{\mu}_c$, and 
obtained a similar phase diagram to that in Fig.\ref{fig:PD1}.

The above interpretation of the phase structure is 
supported by calculating the pseudo-spin correlation functions, 
\be
C_z(r)&=&\frac{1}{L^{3}}\sum_{r_0}\langle{S}^z_{r_0}{S}^z_{r+r_0}\rangle,\nn
C_{xy}(r)&=&\frac{1}{L^{3}}\sum_{r_0} \sum_{\gamma=x,y}
\langle S^\gamma_{r_0}S^\gamma_{r+r_0}\rangle,
\ee
which are used for identification of each phase (see later discussion and
Fig.\ref{fig:spin_AF}).

\begin{figure}[t]
\begin{center}
\includegraphics[width=7cm]{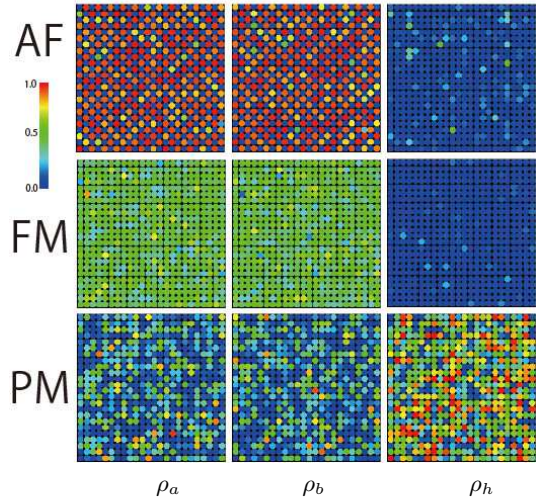} \\
\hspace{1.3cm} $\rho_a$ \hspace{1.4cm} $\rho_b$ \hspace{1.7cm} $\rho_h$
\caption{
Snapshots of three densities $\rho_a$, $\rho_b$, and $\rho_h$ of Eq.(\ref{density}) in a XY plane in three phases of Fig.1 for $L=24$. 
 From the above, $(c_1, \alpha)=(3.0, -0.5)$ (AF phase),  
$(c_1, \alpha)=(3.0, -1.5)$ (FM phase), 
and  $(c_1, \alpha)=(0.5, -0.5)$ (PM phase).}
\label{fig:density1}
\end{center}
\end{figure}

To understand the properties of each phase in an intuitive manner,
it is helpful to examine typical configurations of variables.
In Fig.\ref{fig:density1}, we present snapshots
of three densities, 
\be
\rho_a\equiv \la a^\dag_r a_r\ra,\ \rho_b\equiv \la b^\dag_r b_r\ra,\ 
\rho_h\equiv \la \phi^\dag_r \phi_r\ra,
\label{density}
\ee at each phase.
They are consistent with our previous interpretation of each phase given in
the explanation of Fig.1.
In the AF phase, atoms $a$ and $b$ form the checkerboard configuration.
In the FM state, the both atoms $a$ and $b$ have rather homogeneous density,
and the hole density is very low as the energy dominates over
the entropy at low $T$.
On the other hand, the PM phase has a lower atomic density as the entropy
dominates over the energy at relatively high $T$.

\subsection{Superfluid}

\begin{figure}[t]
\begin{center}
\hspace{-0.3cm}
\includegraphics[width=4.5cm]{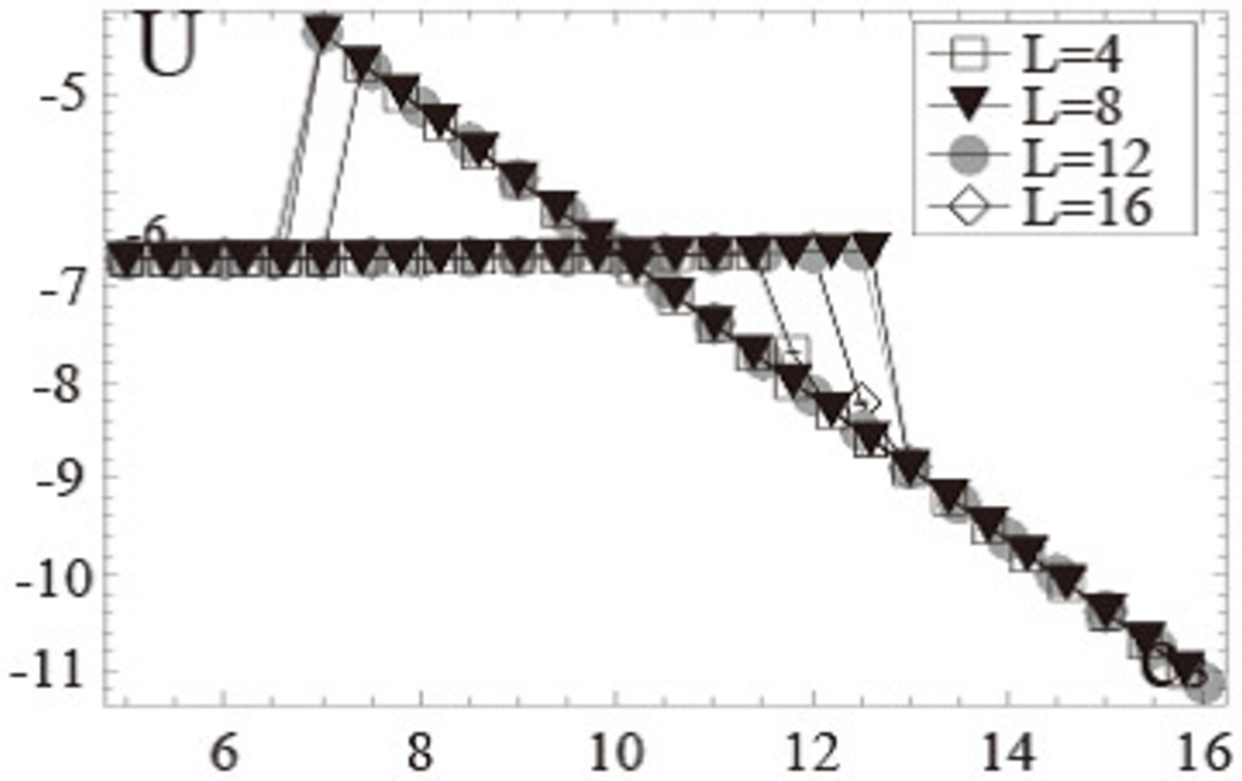}
\includegraphics[width=4.2cm]{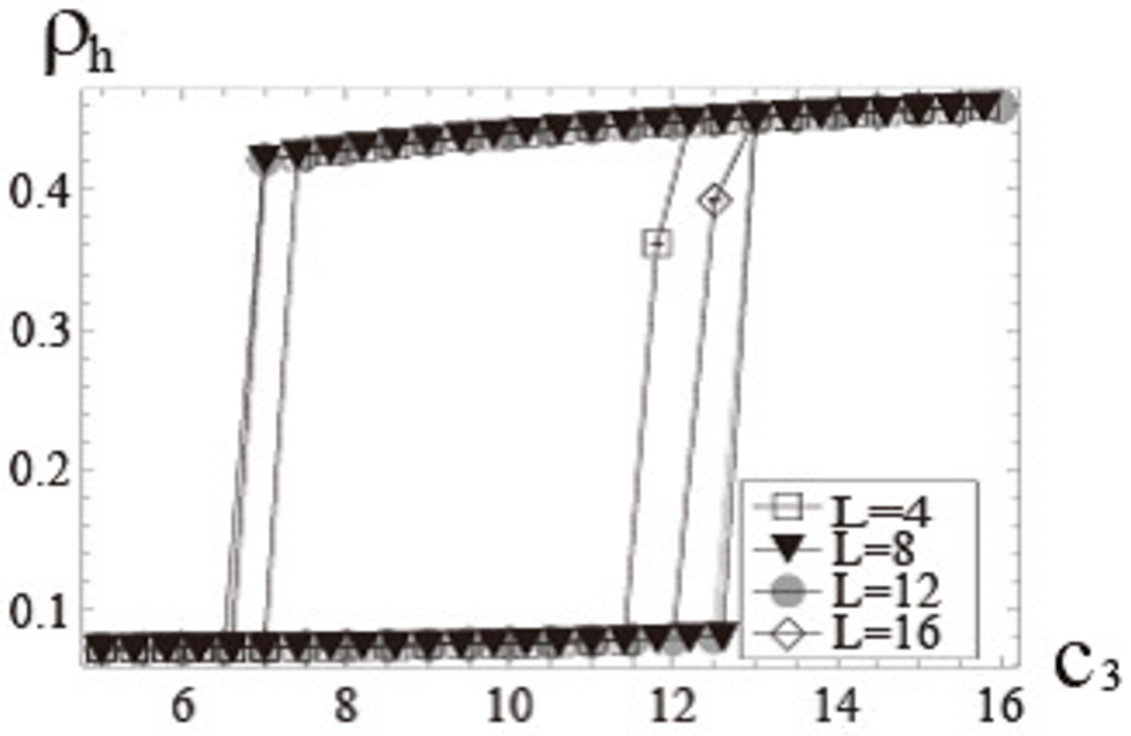}
\hspace{0.1cm}
\includegraphics[width=3.9cm]{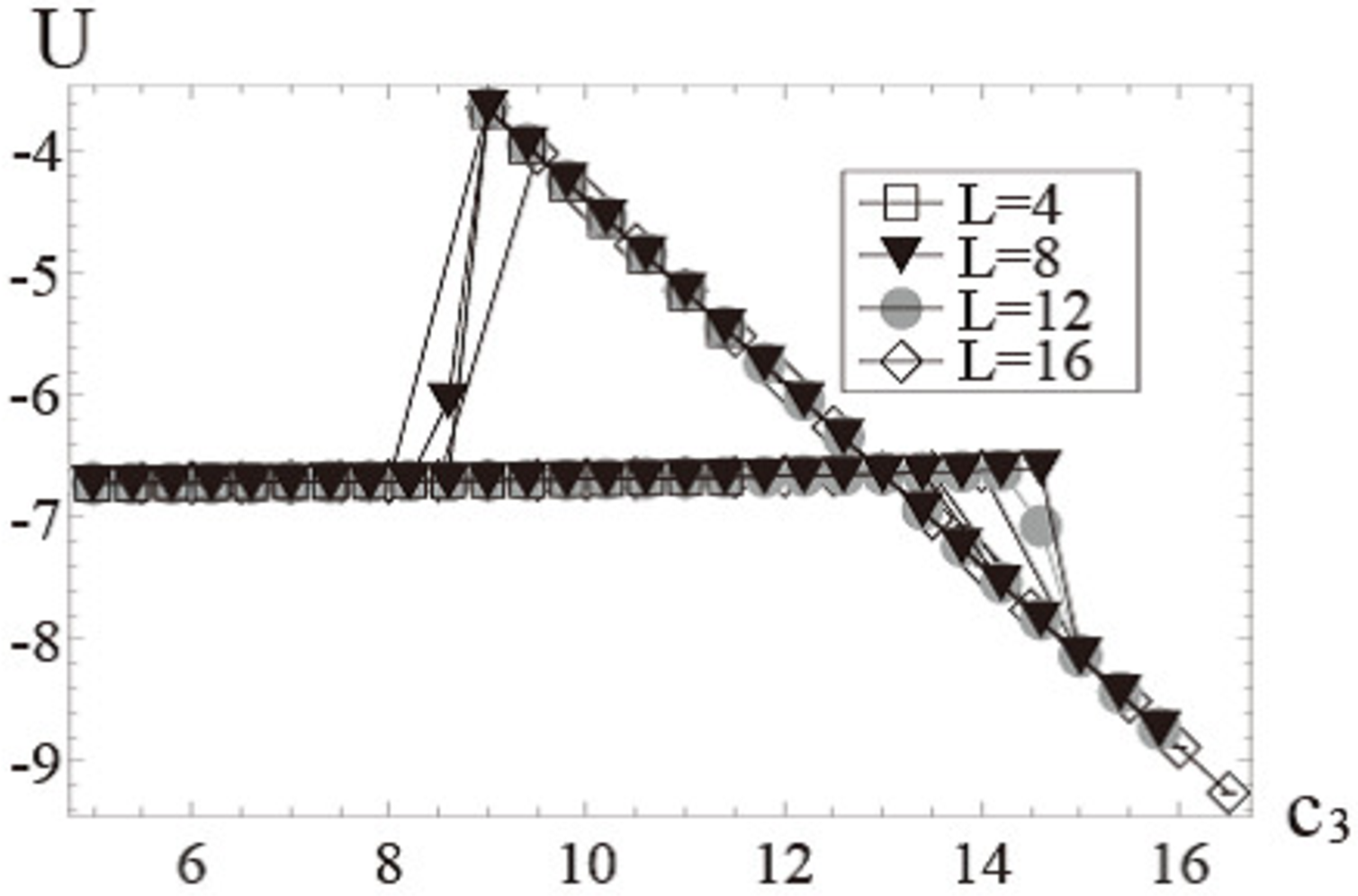}
\hspace{0.2cm}
\includegraphics[width=3.9cm]{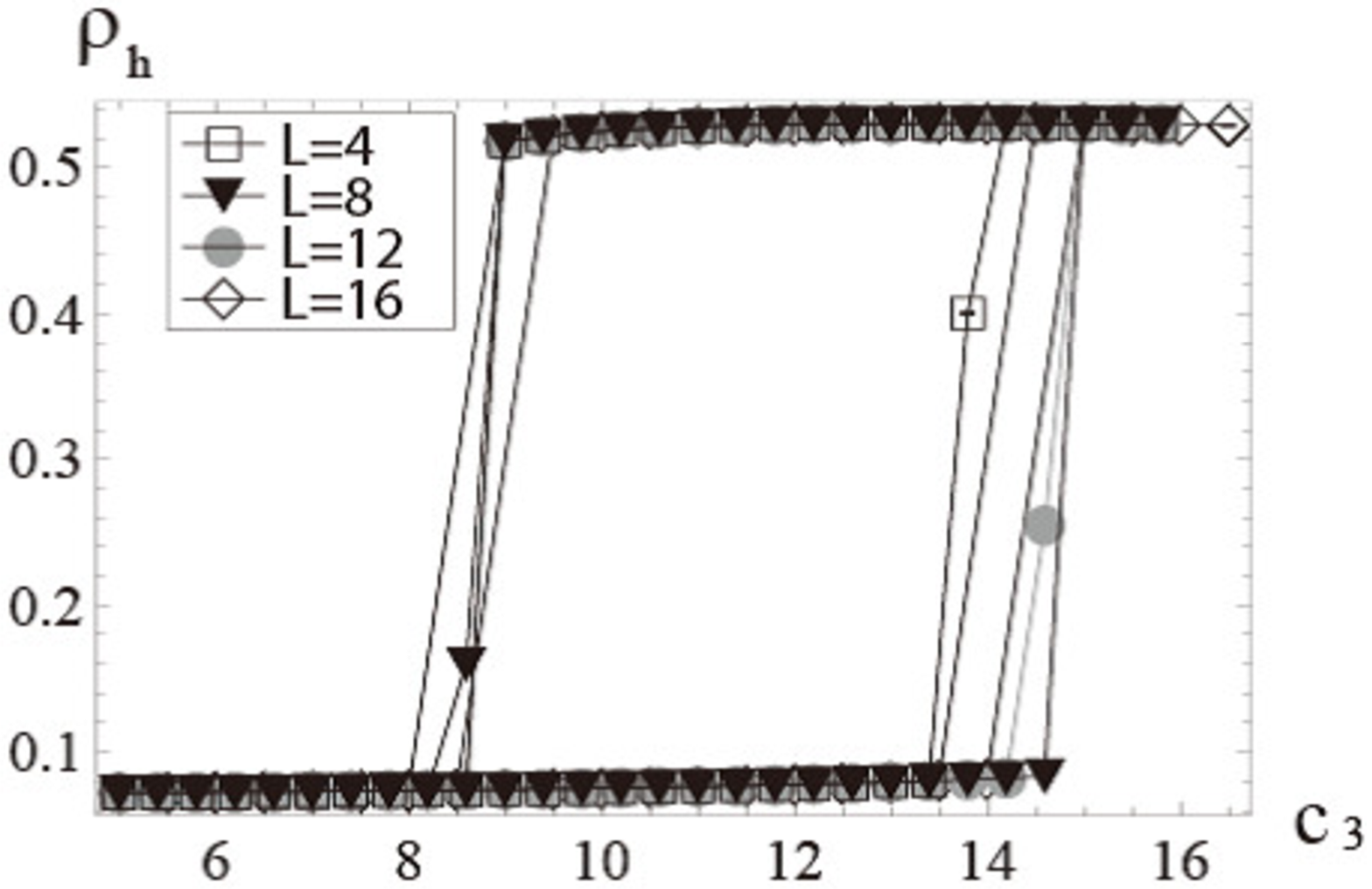}
\caption{
$U$ and the hole density $\rho_h$ as a function of $c_3$
for $\alpha=-0.5$ (upper panels) and $0.5$ (lower panels) with $c_1=3.0$.
Hysteresis loop indicates the transition from AF state to SF
is of first-order.
Similar hysteresis loops are obtained for other values of $\alpha$ and 
$c_1$ for the AF phase.
}
\label{fig:SF1}
\end{center}
\end{figure}

In this subsection, we shall consider the case of finite hopping amplitude $t$.
We verified numerically that the global phase structure of 
Fig.\ref{fig:PD1} remains intact for small $t$ (i.e., $c_3$).
However as $c_3$ is increased, phase transition to
SF state takes place at some critical values 
$c_3=c_{3c}(c_1,\alpha)$.
The transition from the AF phase at $c_3<c_{3c}$ to the SF phase
at $c_3>c_{3c}$ is of strong first order as $U$ and the 
hole density $\rho_h$ in Fig.\ref{fig:SF1} show.
We employed the specific update methods for the MC simulations in order to generate pre-choice configurations efficiently for the first-order phase 
transition\cite{BtJ1,nakano}.
Nevertheless, the obtained
$U$ and the hole density $\rho_h$ 
exhibit large hysteresis loops as $c_3$ varies.

\begin{figure}[b]
\begin{center}
\includegraphics[width=5cm]{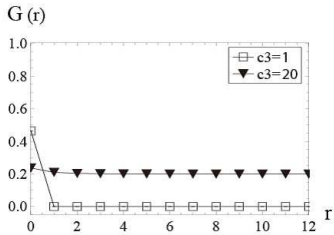}
\caption{
Boson correlation function $G(r)=G_a(r)=G_b(r)$ for 
$c_1=3.0$, $\alpha=0.5$.
Finite LRO for $c_3=20$ indicates that the phase transition in 
Fig.\ref{fig:SF1} is a SF transition.
}
\label{fig:boson_AF}
\end{center}
\end{figure}
\begin{figure}[t]
\begin{center}
\includegraphics[width=4cm]{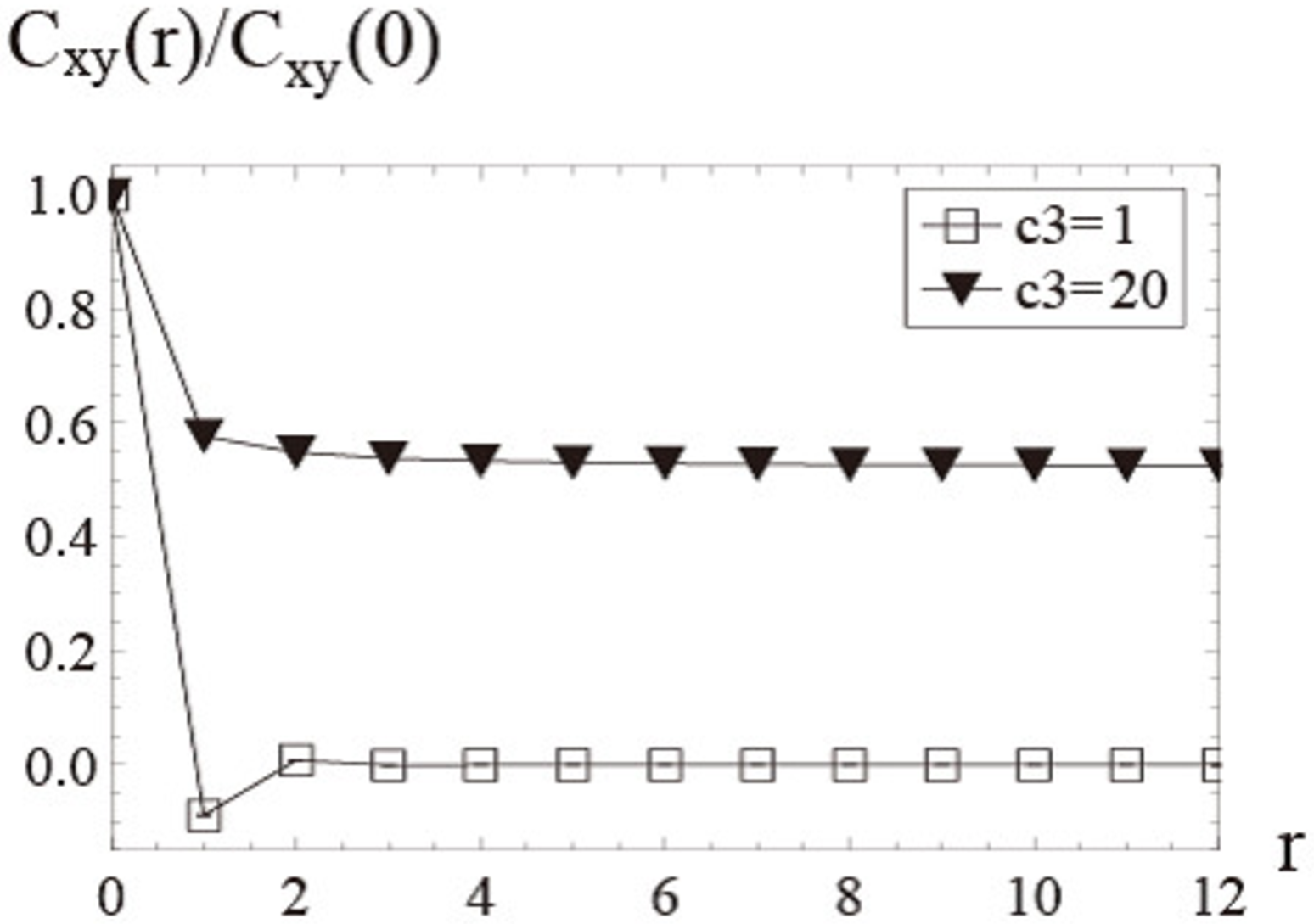}
\includegraphics[width=4cm]{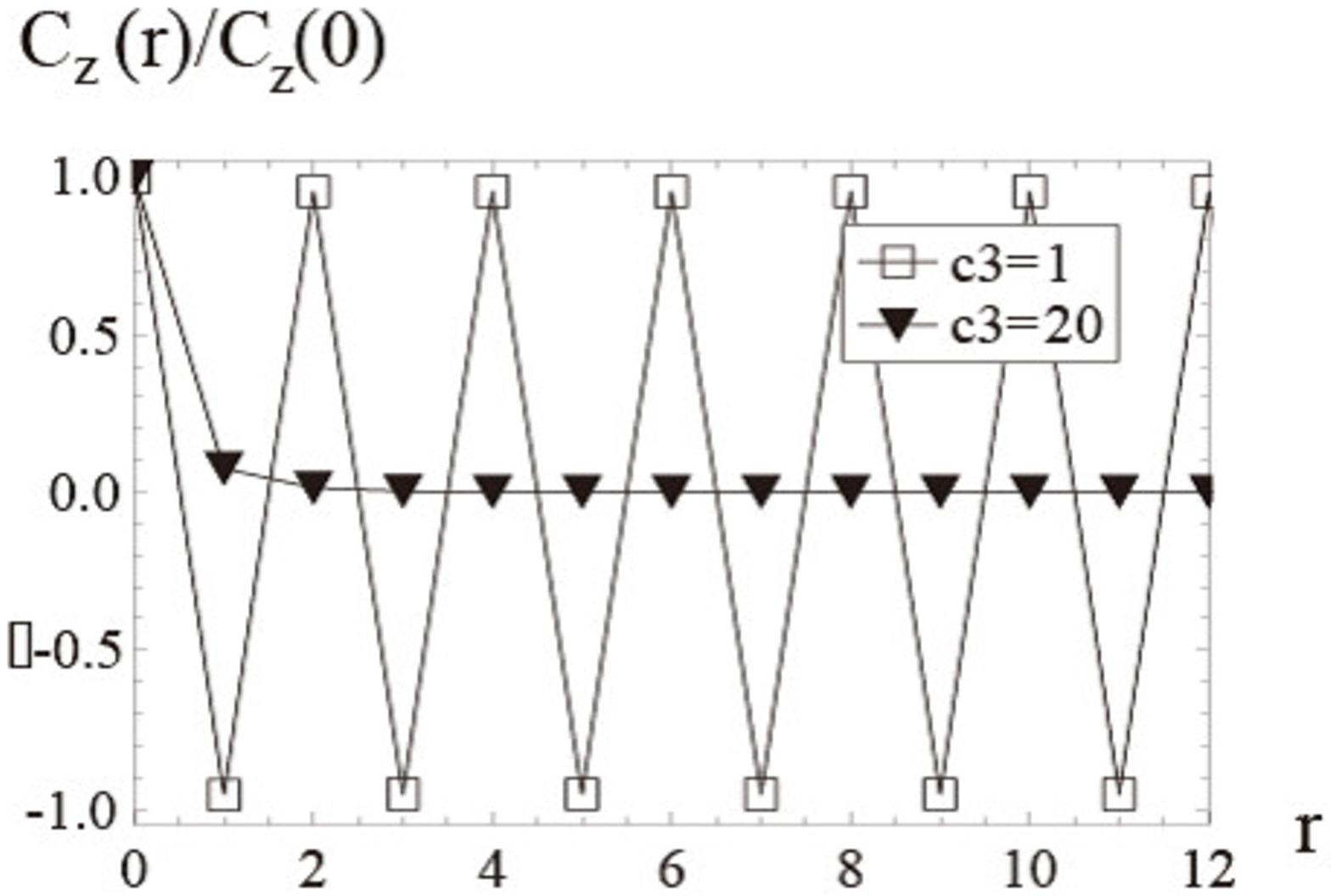} \\
\includegraphics[width=4cm]{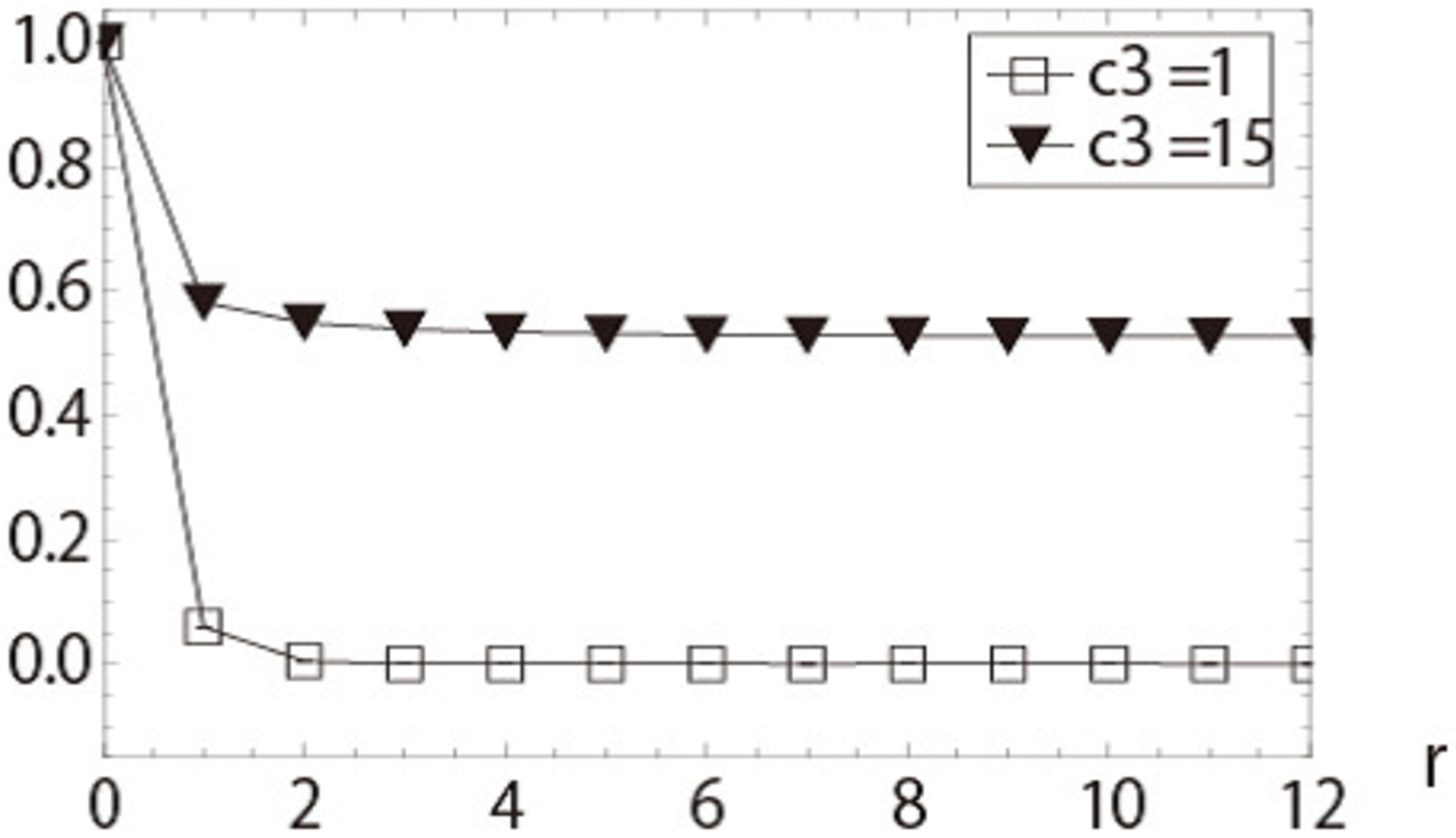}
\includegraphics[width=4cm]{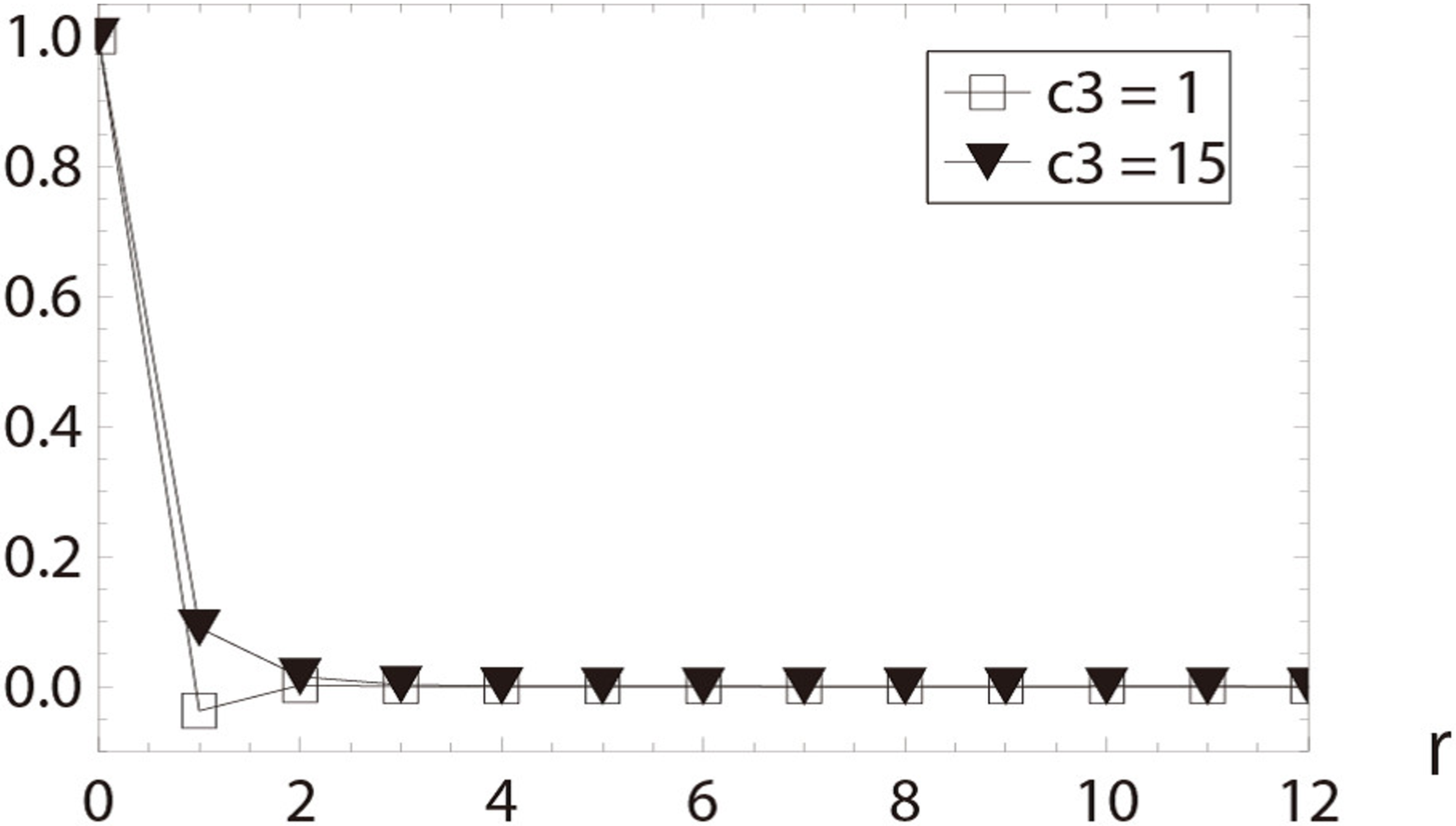} \\
\includegraphics[width=4cm]{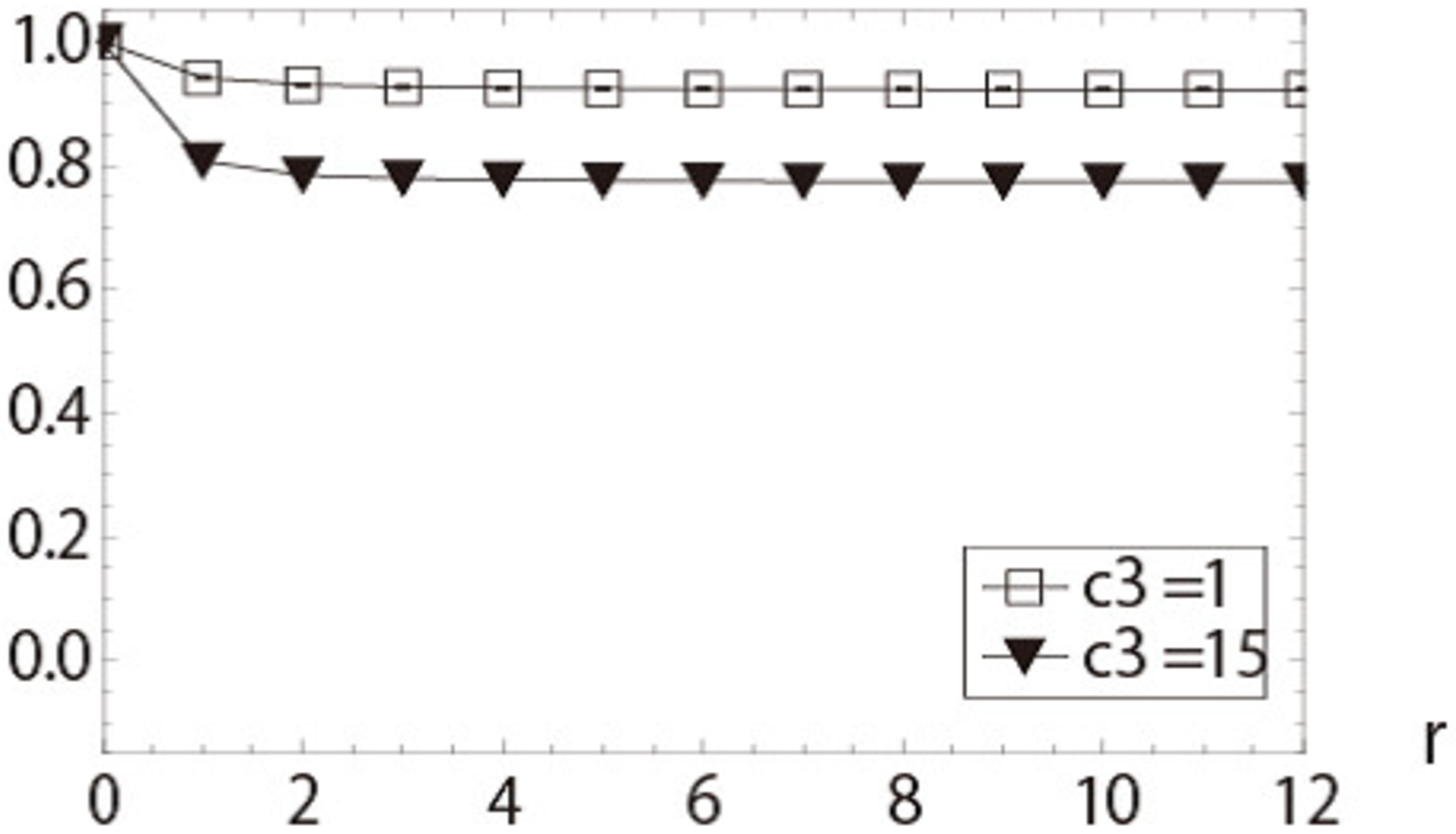}
\includegraphics[width=4.5cm]{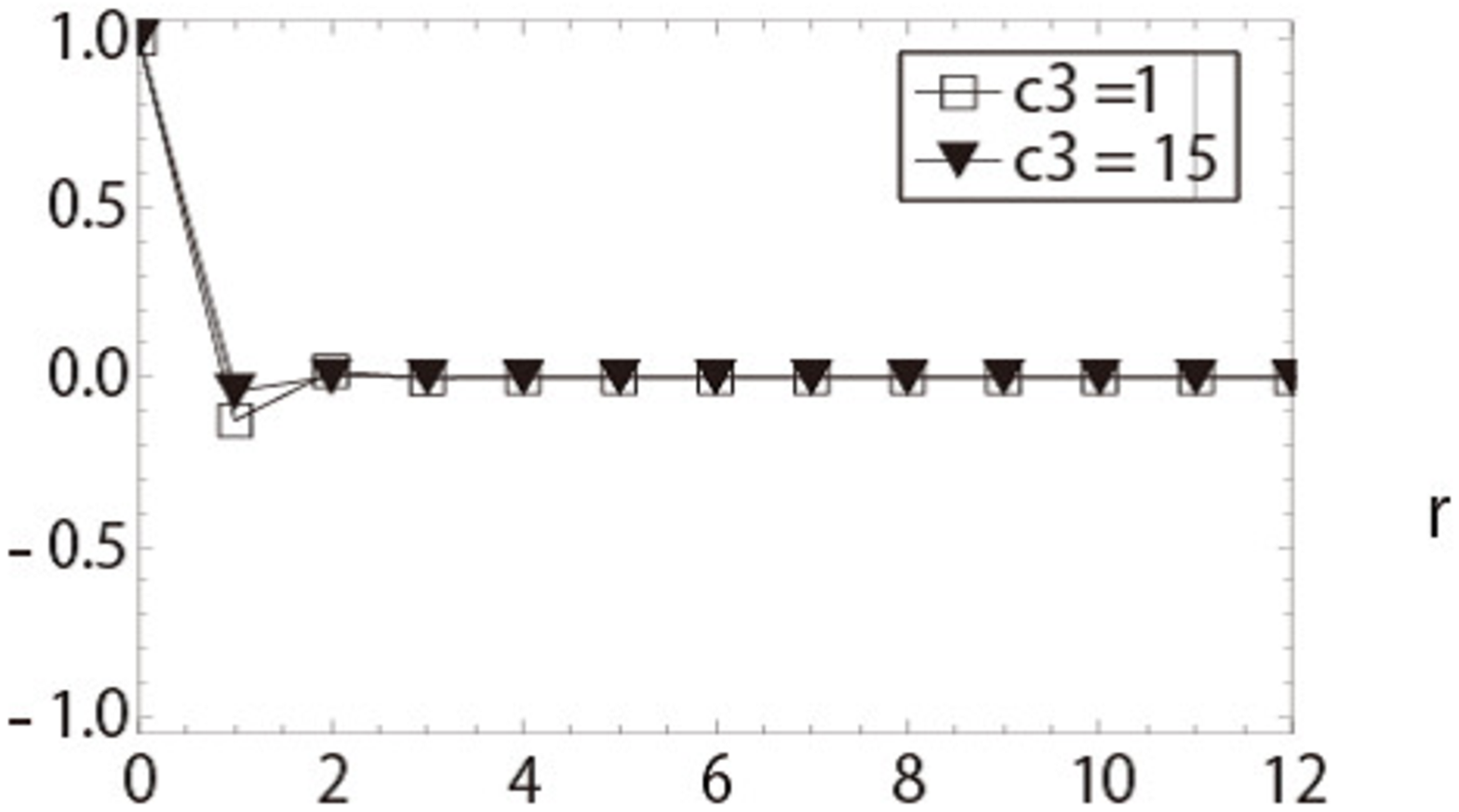}
\caption{
(Top) Pseudo-spin correlation functions $C_{xy}(r)$ and $C_z(r)$
in the AF and SF states for $\alpha=0.5$ and $c_1=3.0$.
In SF state, a FM LRO appears in the $xy$-plane of 
the pseudo-spin space.
(Middle) $C_{xy}(r)$ and $C_z(r)$ in the PM and SF states for $\alpha=-1.5$
and $c_1=0.5$.
(Bottom) $C_{xy}(r)$ and $C_z(r)$ in the FM and SF states for $\alpha=-1.5$
and $c_1=3.0$.
}
\label{fig:spin_AF}
\end{center}
\end{figure}

To verify that the SF state is realized for $c_3>c_{3c}$, we calculated 
the boson correlation function $G_a(r)$ and  $G_b(r)$,
\begin{eqnarray}
G_a(r)&=&{1 \over L^3}\sum_{r_0}\langle \phi^\dagger_{r_0}z_{1,r_0}
z^\dagger_{1,r_0+r}\phi_{r_0+r}\rangle,  \nonumber  \\
G_b(r)&=&{1 \over L^3}\sum_{r_0}\langle \phi^\dagger_{r_0}z_{2,r_0}
z^\dagger_{2,r_0+r}\phi_{r_0+r}\rangle,
\end{eqnarray}
and if $G_a(r), \; G_b(r) \rightarrow$ finite as $r\rightarrow \infty$, the SF is realized.
The results are shown in Fig.\ref{fig:boson_AF}.
It is obvious that $G_a(r)=G_b(r)$ in the present case, and it has a nonvanishing
LRO for $c_3=20$ indicating existence of a finite density of SF.
In Fig.\ref{fig:spin_AF} we also show the calculation of 
the pseudo-spin correlation functions 
$C_z(r)$ and  $C_{xy}(r)$.
The results show that the phase transition to the SF state accompanies a
transition from the Ising-like AF LRO to the XY-FM LRO\cite{FMSF}.
This result is in sharp contrast with the result obtained by the MFT.
The present numerical study indicates that the SS phase 
predicted in MFT, in which the AF LRO and SF coexist, 
does not appear in the present model. 
As the phase transition to the SF phase takes place at
$c_{3}=\beta t\sim O(10)$ and $c_1 [\sim O({\beta t^2 \over U},
{\beta t^2 \over V})]=3$, the critical region is located at  
$t/U, \ t/V \sim 3/10 \ll 1$ in the original Hubbard model.
Therefore the above obtained results in the $t$-$J$ model are also
applicable for the bosonic Hubbard model.

\begin{figure}[b]
\begin{center}
\includegraphics[width=7cm]{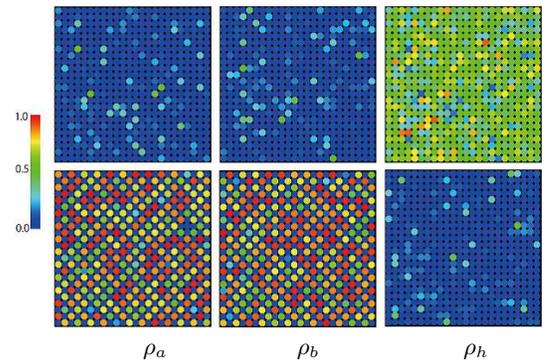} \\
$\hspace{1cm} \rho_a$ \hspace{1.5cm} $\rho_b$ \hspace{1.7cm} $\rho_h$
\caption{
Snapshots of three densities $\rho_a$, $\rho_b$, and $\rho_h$ 
of Eq.(\ref{density}) in a XY plane
for $L=24, c_1=3.0, c_3=10.0, 
\mu=0.0, \alpha=0.5$. The upper ones are in the FM+SF phase with 
the higher energy $U/N=-4.39$
and the lower ones are in the AF phase with the lower energy $U/N= -6.68$.
}
\label{fig:density2}
\end{center}
\end{figure}

\begin{figure}[t]
\begin{center}
\includegraphics[width=7cm]{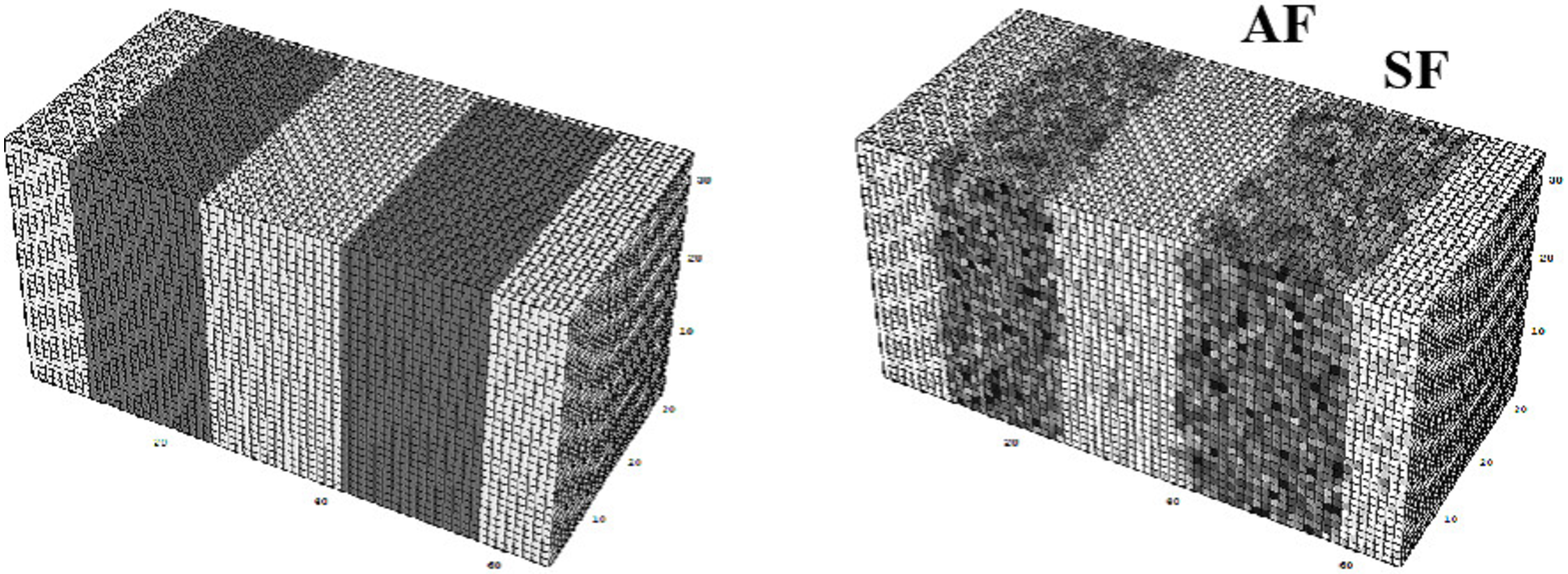}  \\
\vspace{0.5cm}
\includegraphics[width=8cm]{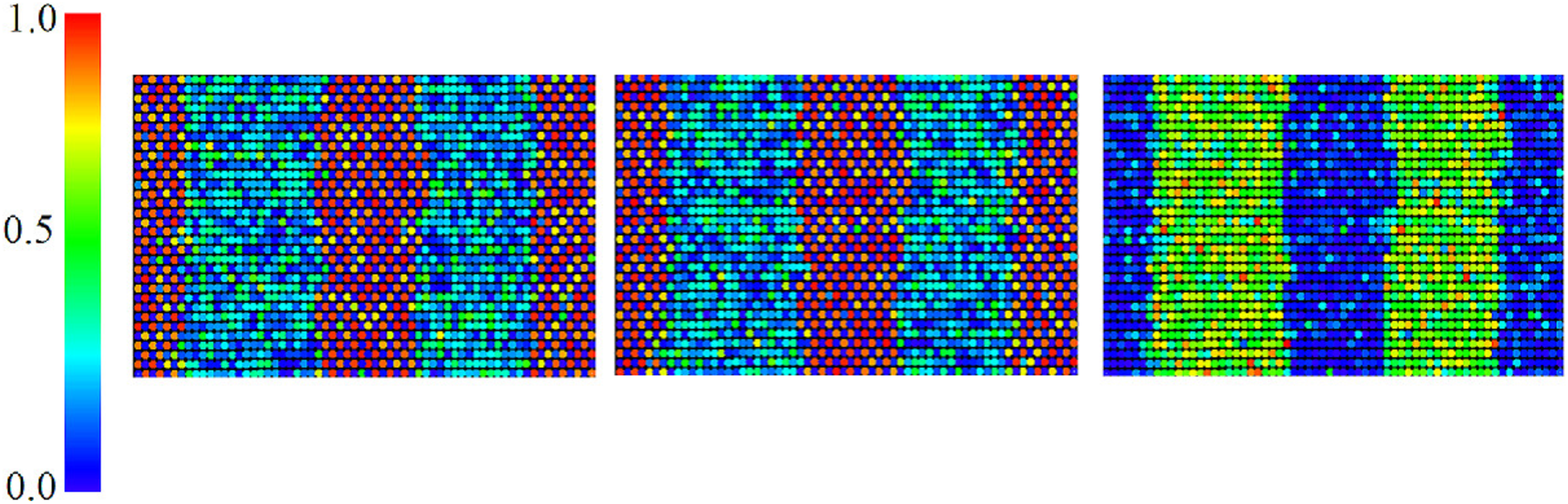}  \\
\vspace{-0.5cm}
$\hspace{1cm} \rho_a$ \hspace{1.5cm} $\rho_b$ \hspace{1.7cm} $\rho_h$
\caption{
(Upper panels) (Left) Initial configuration with phase separation and 
(Right) configuration after $60\times 10^5$ sweeps at the critical point.
Dark regions represent SF with relatively high hole density and bright regions
represent AF solid. 
(Lower panels) Densities $\rho_a$, $\rho_b$ and $\rho_h$ in a horizontal plane
in configuration after $60\times 10^5$ sweeps.
}
\label{PS}
\end{center}
\end{figure}

There are two kinds of SF, one made of atom $a_r$ and the other made of $b_r$.
It is interesting to see how each SF
behaves in the hysteresis region of the first-order phase transition.
In Fig.\ref{fig:density2} we present snapshots of typical configurations of $\rho_a$
and $\rho_b$ for $c_3=10.0$ to check the possibility that the AF solid and SF exist 
separately in every state of update.
We found that on the $c_3$-decreasing line of the hysteresis loop 
the pure FM+SF state
is realized, whereas the pure AF state exists on the $c_3$-increasing line.
This indicates that in real experiments there exists a genuine phase 
transition point in the middle of the hysteresis loop in the MC simulation and 
the internal energy has a sharp discontinuity at that transition point.
At the discontinuity point, immiscible state of the AF solid and SF is realized.
In order to verify this expectation, we performed MC simulation starting with
a half-AF and half-SF configuration and searched a ``genuine critical coupling" 
$c_{3c}$ at which this phase-separated configuration is stable during
MC update.
For $\alpha=0.5$ and $c_1=3.0$, we found $c_{3c}=11.36$, see Fig.\ref{PS}.
This result, which shows that the phase separation takes place in the present 
3D system, is consistent with the result of the previous study on the system
at $T=0$\cite{BtJ2}.

\begin{figure}[t]
\begin{center}
\includegraphics[width=4cm]{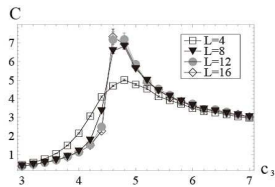}
\caption{
Transition from PM to FM+SC.
Specific heat for $c_1=0.5$ and $\alpha=-1.5$ vs $c_3$.
The peak of $C$ has a systematic $L$ dependence
of a typical second-order transition.  
}
\label{PMSF}
\end{center}
\end{figure}
\begin{figure}[t]
\begin{center}
\includegraphics[width=4cm]{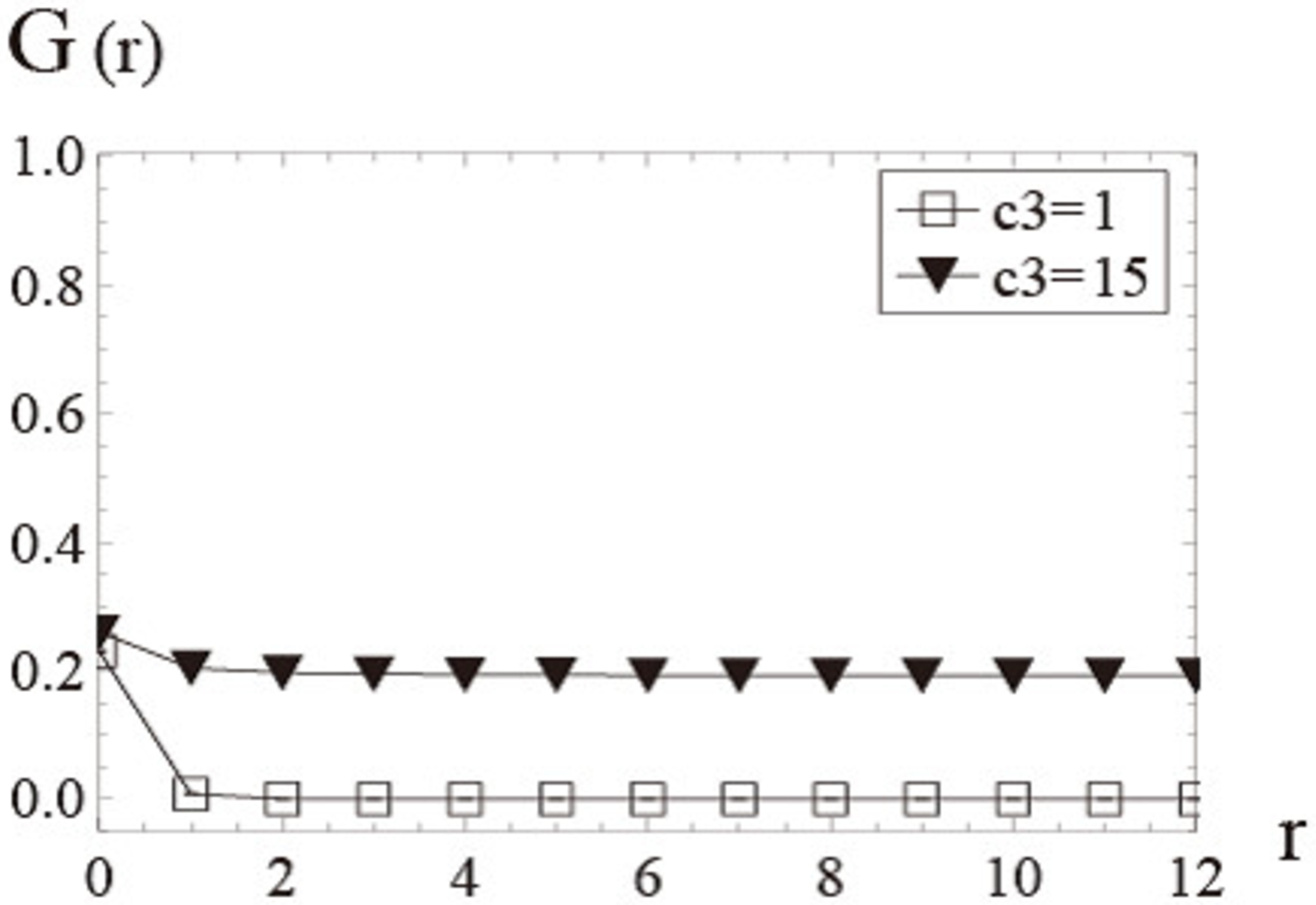}
\includegraphics[width=4cm]{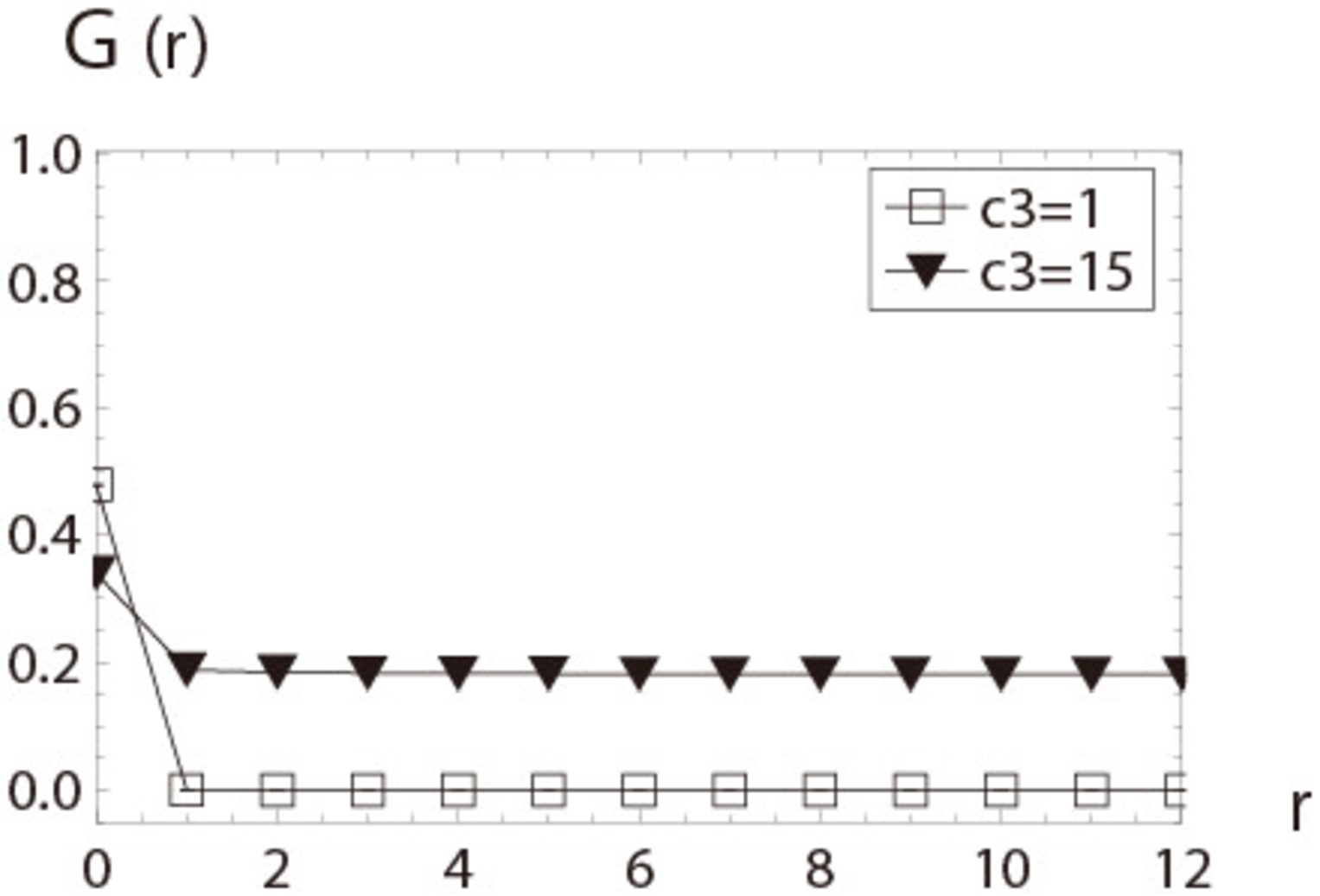}
\vspace{-0.3cm} \\
\hspace{0.5cm} (a) \hspace{3.5cm} (b)
\caption{
Boson correlation function $G(r)=G_a(r)=G_b(r)$ for 
$c_1=0.5$, $\alpha=-1.5$ (Left) and
$c_1=3$, $\alpha=-1.5$ (Right).
Finite LRO for $c_3=15$ indicates that the phase transition in 
Figs.\ref{PMSF} and \ref{fig:FMSFC} is a SF transition.}
\label{fig:boson_PMFM}
\end{center}
\end{figure}
\begin{figure}[t]
\begin{center}
\includegraphics[width=4cm]{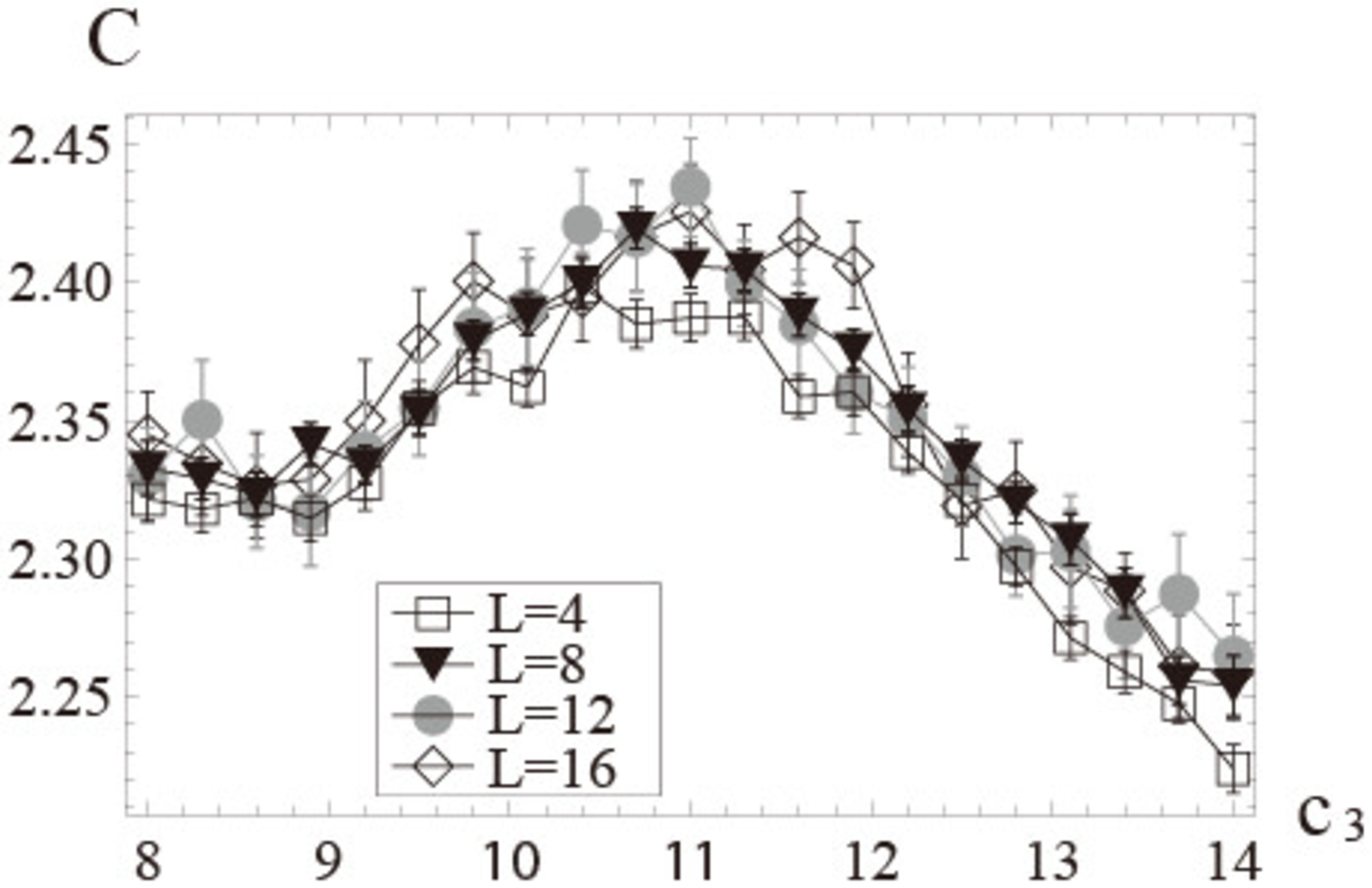} \\
\begin{minipage}{0.49\hsize}
\includegraphics[width=4cm]{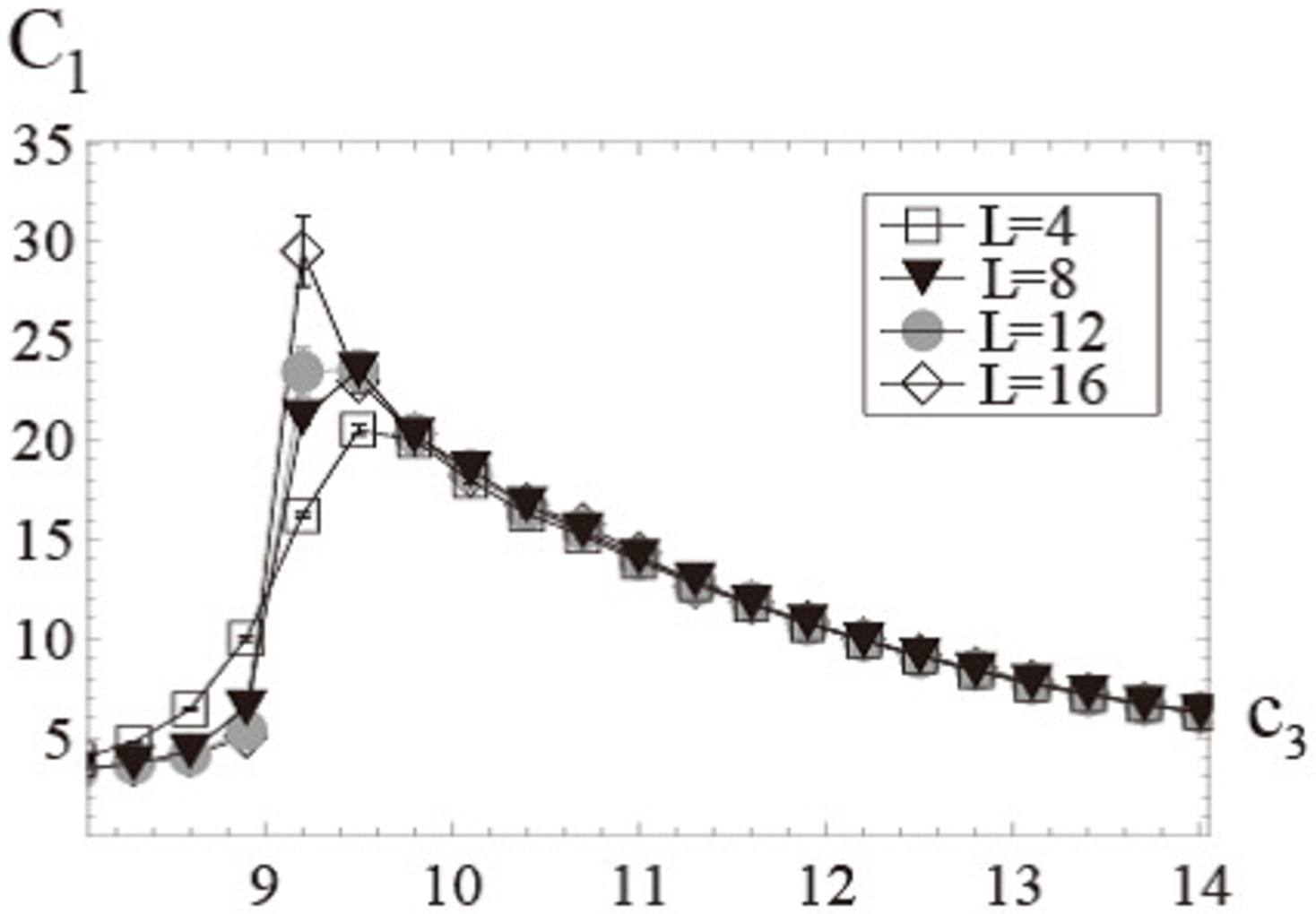} 
\end{minipage}
\begin{minipage}{0.49\hsize}
\vspace{-0.2cm}
\includegraphics[width=3.5cm]{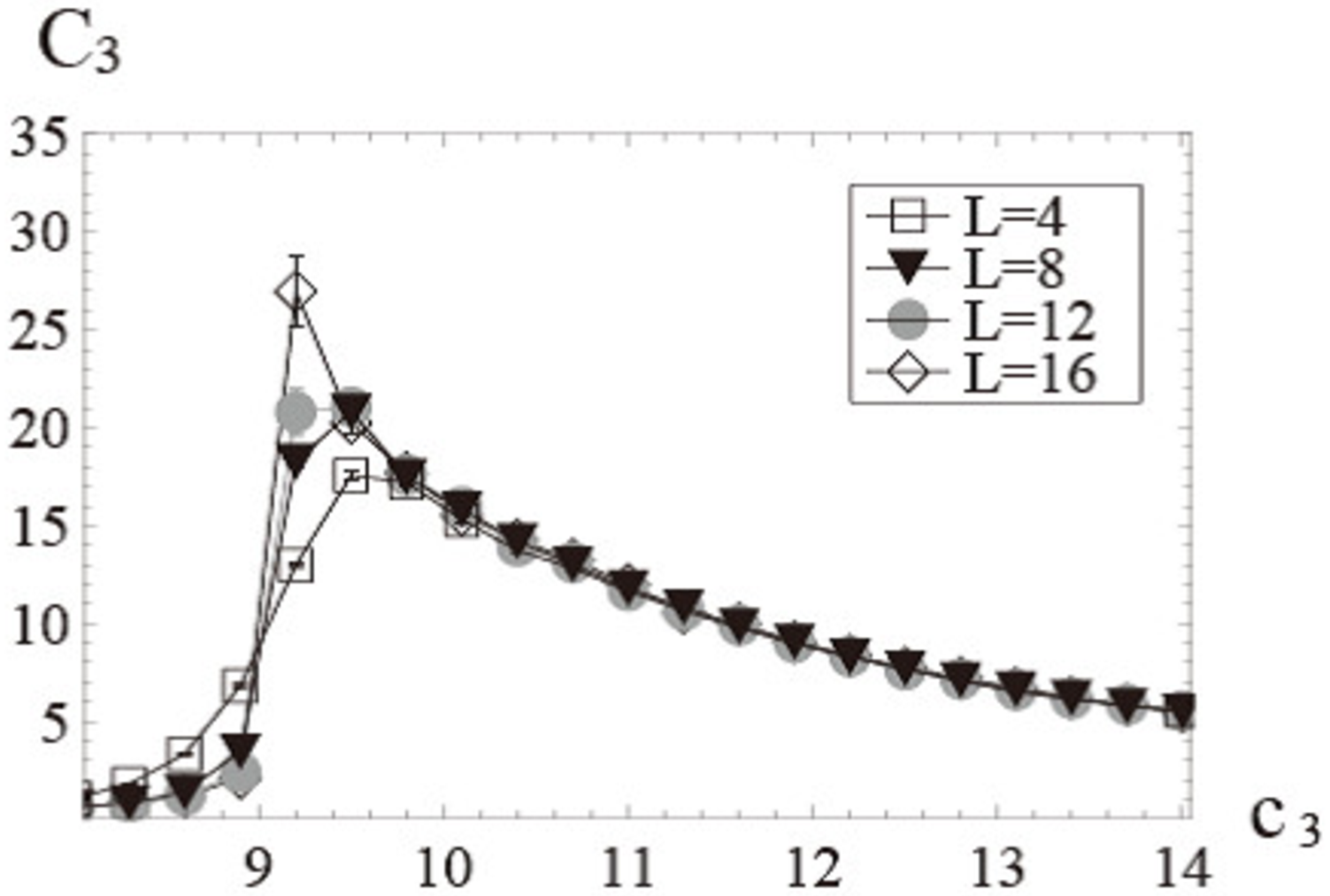}
\end{minipage}
\caption{
Specific heat of phase transition from FM to FM+SC,
$c_1=3.0$ and $\alpha=-1.5$ exhibits rather peculiar behavior,
whereas specific heat of each term $c_1$, $c_3$ shows typical system-size
dependence of second-order phase transition. 
}
\label{fig:FMSFC}
\end{center}
\end{figure}

Let us turn to the PM $\rightarrow$ SF transition. 
In Fig.\ref{PMSF} we present $C$ for $c_1=0.5$ and $\alpha=-1.5$.
$C$ exhibits a sharp peak at $c_3\simeq 4.6$, which indicates existence of a
second-order phase transition.
We calculated the boson correlation function and verified that a SF appears
for $c_3>4.6$.
See Fig.\ref{fig:boson_PMFM}a.

We also verified that a transition from the FM to FM+SC takes place 
as $c_3$ is increased.
In the critical region, the total specific heat $C$ exhibits
rather peculiar behavior, but the ``specific heat" of each term, 
defined by $c_i \equiv \langle (E_i-\langle E_i \rangle )^2\rangle/L^3$
for each term $E_i$ in the Hamiltonian,
shown in Fig.\ref{fig:FMSFC} exhibits
typical behavior of the second-order phase transition.
In Fig.\ref{fig:boson_PMFM}b, we show the boson correlation function
for Fig.\ref{fig:FMSFC}.
The existence of the finite LRO means that the phase transition in 
Fig.\ref{fig:FMSFC} is again a transition to SF.

\begin{figure}[t]
\begin{center}
\includegraphics[width=4cm]{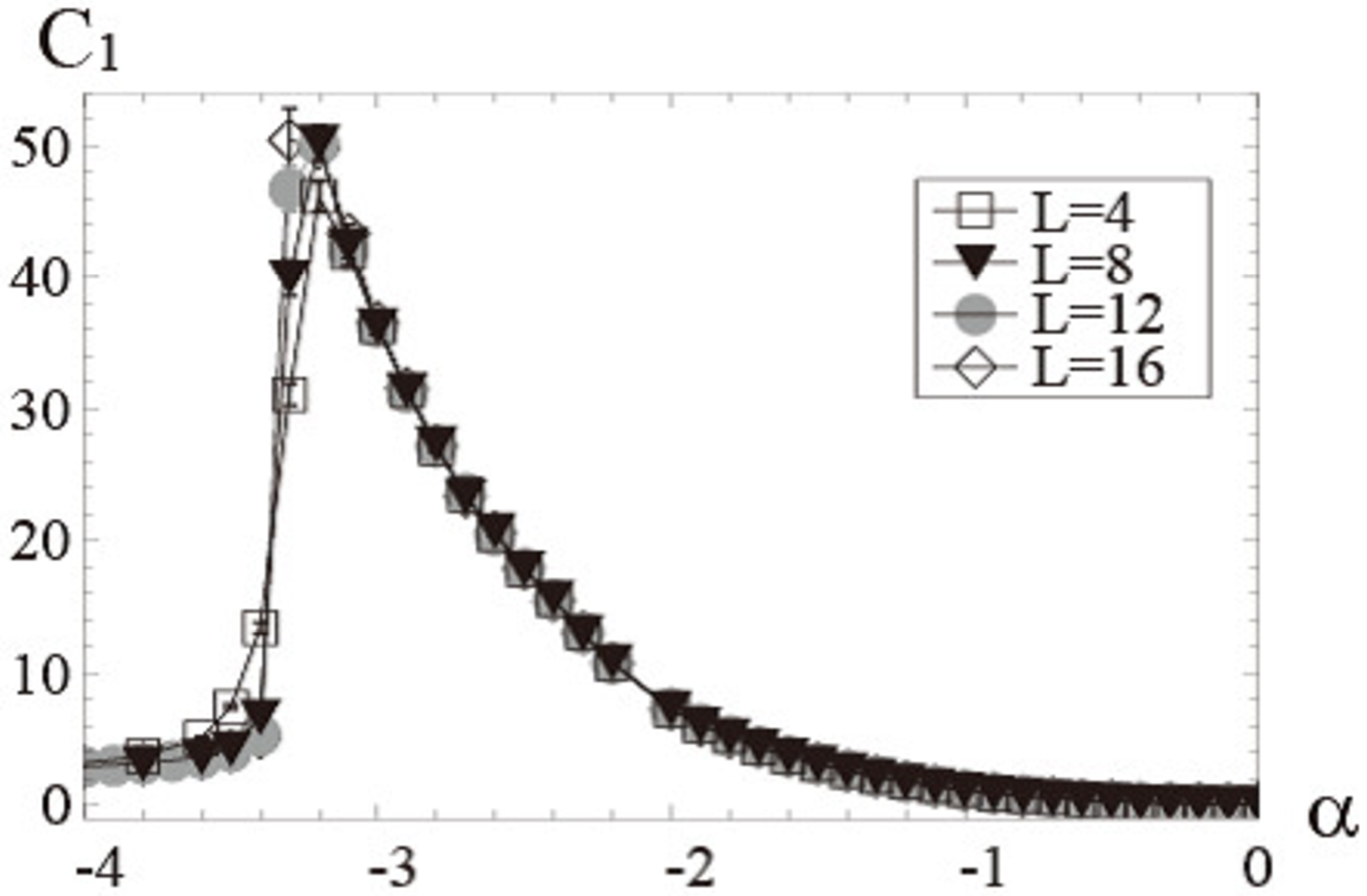}
\includegraphics[width=4cm]{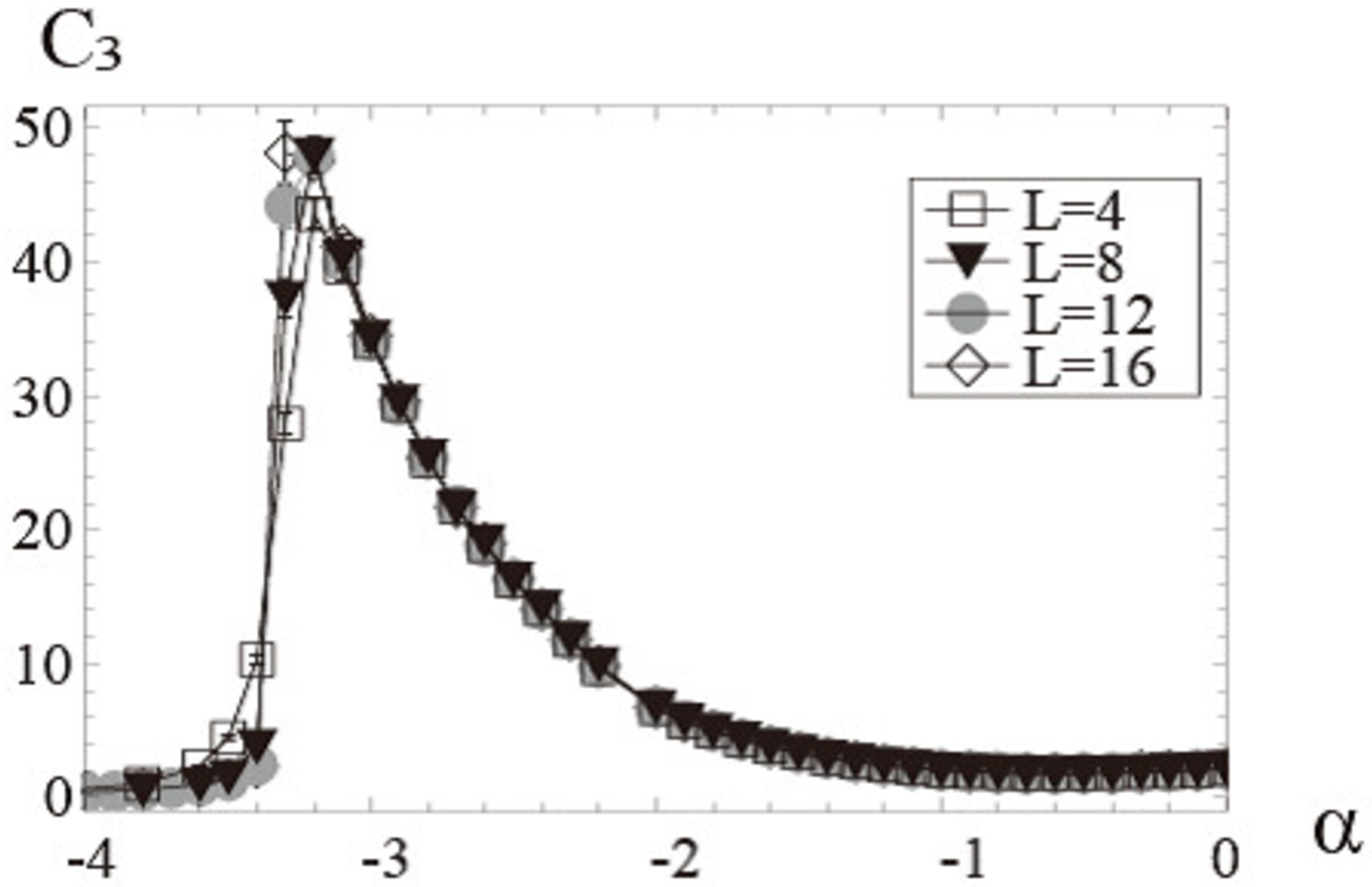}
\includegraphics[width=4cm]{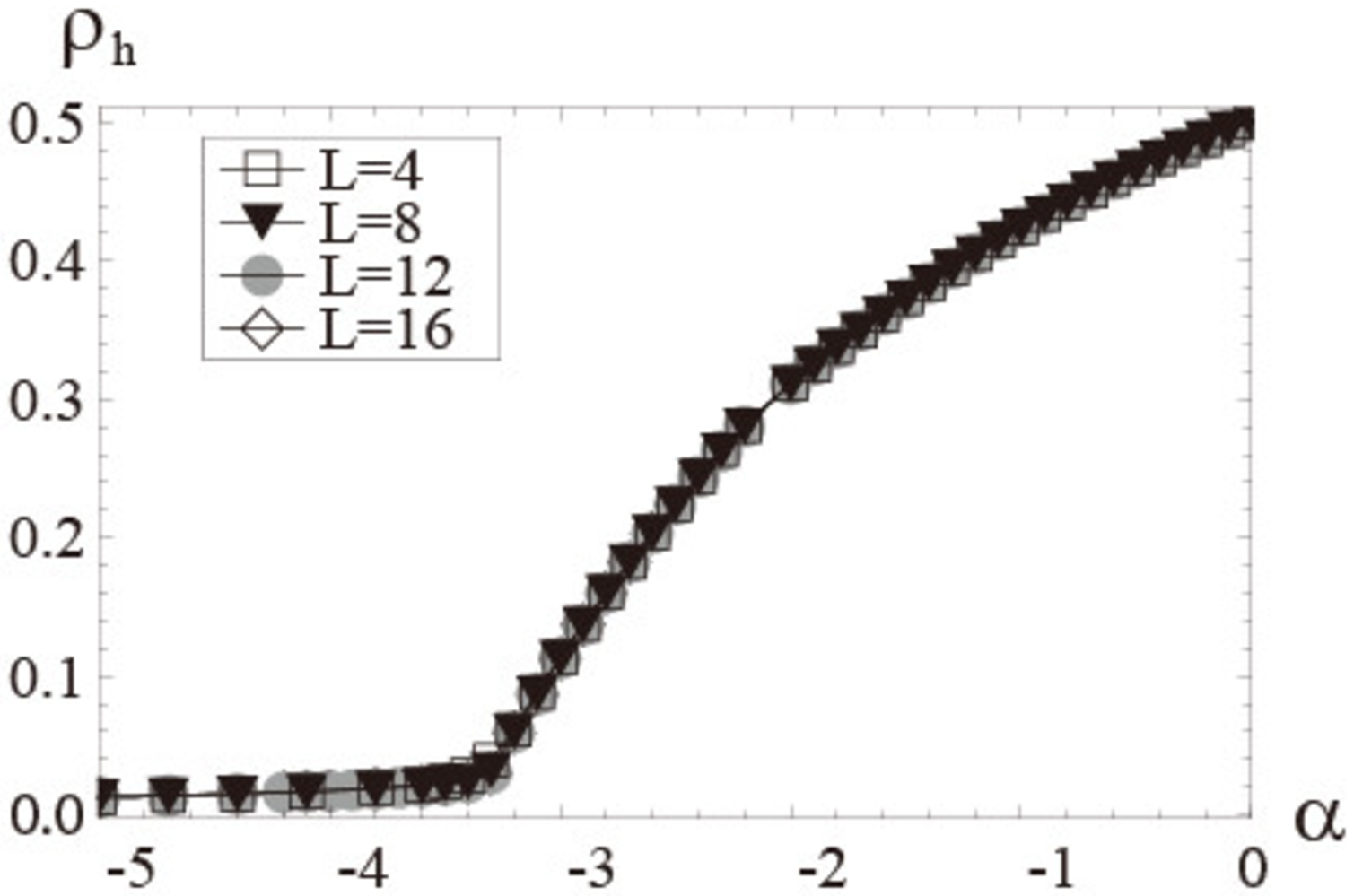} 
\includegraphics[width=4cm]{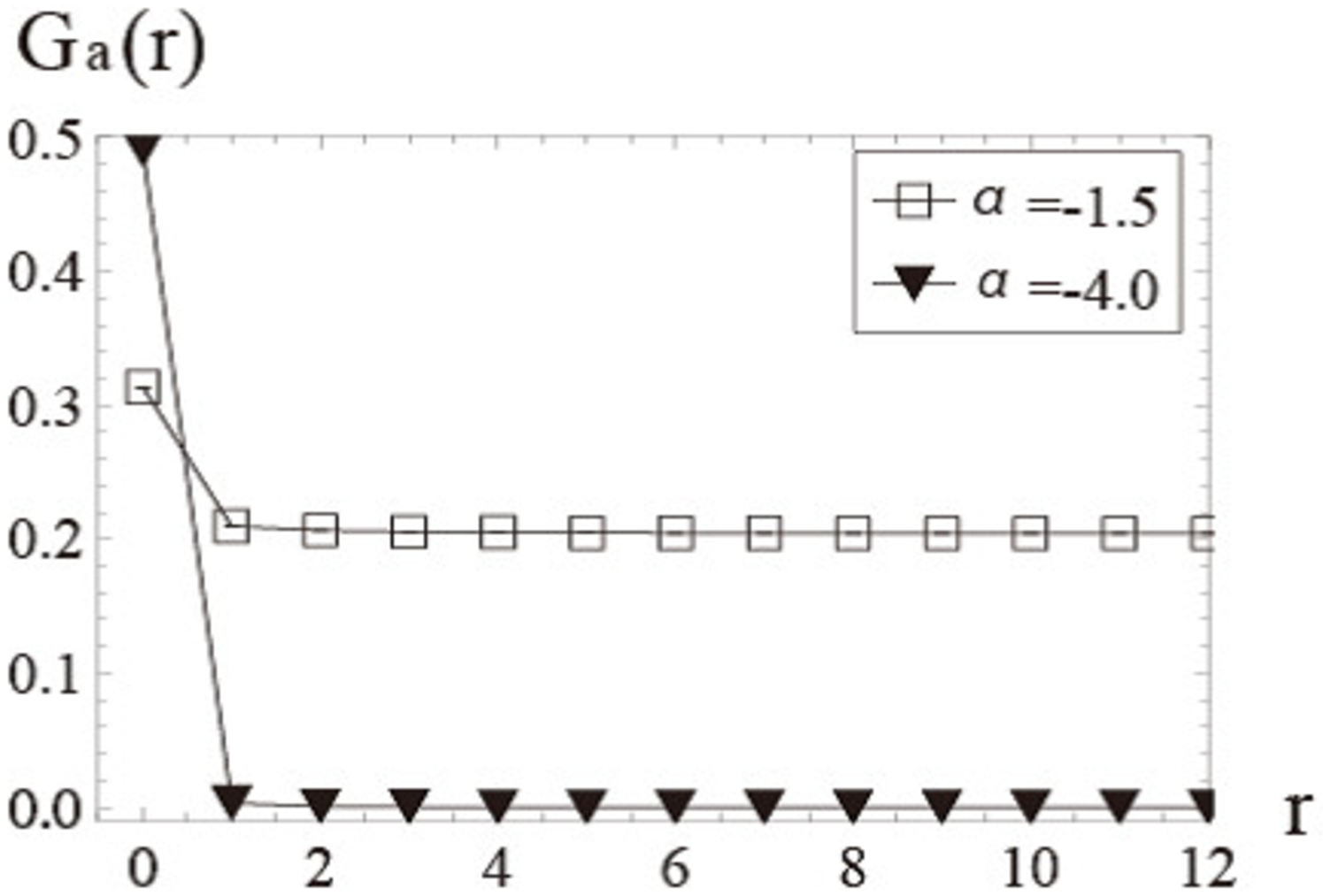} 
\caption{
Specific heat of each term as a function of $\alpha$ for
$c_1=3.0$ and $c_3=20$. 
There exist no anomalous behavior at $\alpha \simeq -1$.
Peaks in the specific heats at $\alpha\simeq -3.2$ mean a phase 
transition to vanishing SF, for hole density is vanishingly small and 
boson correlation vanishes for $\alpha < -3.2$.
}
\label{fig:FMSFs}
\end{center}
\end{figure}

We observed that all three phases at $t=0$, i.e., PM, AF and FM phases,
evolve into the SF state as $t$ is increased.
Then, 
it is quite interesting to see if there is a phase boundary between these SF's
for sufficiently large $c_3$
though all of three phases belong to the FM$+$SF phase.
This problem is closely related with recent experiment investigating two species
SF\cite{2BEC}.
This experiment observed that (im)miscibility of two SF's depends on the
inter and intra-interactions between atoms. 
In Fig.\ref{fig:FMSFs},
we show the specific heat of each term and particle density as a function 
of $\alpha$ for $c_1=3.0$ and $c_3=20$. 
On may expect that there is a phase boundary separating two FM$+$SF phases at 
$\alpha\simeq -1$, but the result exhibits no anomalous behaviors there.
On the other hand, the peaks at $\alpha\simeq -3.2$ accompanies
abrupt decrease of the hole density.
This indicates that there exists a phase transition and that is a transition
into the vanishing SF.
$G_a(r)$ and $G_b(r)$ in Fig.\ref{fig:FMSFs} support this interpretation
because they have no RLO at $\alpha=-4.0$.
The phase $\alpha<-3.2$ is a pure FM state without holes.
We also studied whether phase transition between two SF's takes place as 
the value of $c_1$ is varied, but
we found a similar result to the above as varying $\alpha$, i.e.,
there exists no phase transition between two SF's.

\section{Conclusion}
\setcounter{equation}{0} 

In present paper, we studied the $t$-$J$ model of
two-component hard-core bosons by means of MC simulations.
We considered the system with filling factor up to unity,
and obtained the global phase diagram in the grand-canonical ensemble (GCE).
At vanishing hopping amplitude, there are three phases in the $\alpha-c_1$
plane, PM, AF and FM phases.
As the hopping amplitude is increased, all three phases evolve into SF state
with BEC of atoms.
These obtained results are globally consistent with those for the case of
integer fillings obtained by MFT-type approximation and 
numerical methods\cite{altman,soyler}. 
However, we verified that the SS state, which is predicted to appear by the
MFT, does not exist in the present model in the GCE.
On the other hand, we found that the phase separation of the AF solid and 
SF is realized at the phase transition point.

We also studied if there exists phase boundary between the SF's.
However there are no phase boundaries between them.

Results obtained in the present paper show that the bosonic $t$-$J$ model has 
a very rich phase structure.
We studied the system in the GCE.
It is quite interesting to study the bosonic $t$-$J$ model in the canonical
ensemble with fixed average atomic number.
In particular, an inhomogeneous state may appear near the first-order
phase transition point from the AF solid to the SF.
This problem is under study and results will be reported in a future 
publication.

\bigskip

\acknowledgments 
This work was partially supported by Grant-in-Aid
for Scientific Research from Japan Society for the 
Promotion of Science under Grant No.20540264
and No23540301.



\begin{references}
\bibitem{FMSF}Origin of the XY-FM LRO in two-component SF was explained
in the previous paper Ref.\cite{BtJ1}.
\end{references}
\end{document}